\documentclass[10pt,final,twocolumn,journal]{IEEEtran}

\usepackage{latexsym,amsfonts,setspace}
\usepackage{amssymb,amsmath}
\usepackage{float}
\usepackage{cite}
\usepackage{longtable} 
\usepackage{adjustbox}
\usepackage[ruled,vlined,boxed, linesnumbered]{algorithm2e}

\usepackage{tikz}
\usetikzlibrary{arrows}
\usepackage{tikz-dependency}
\usepackage{mathtools}
\usepackage[T1]{fontenc}
\usepackage{pgfplots}
\usepackage{tabularx}
\usepackage{url}

\makeatletter
\renewcommand{\boxed}[1]{
\tikz[baseline={([yshift=-1ex]current bounding box.center)}] \node [rectangle, minimum width=1ex,rounded corners,draw] {\normalcolor\m@th$\displaystyle#1$};}
\makeatother
\tikzstyle{block} = [rectangle, draw,
    text width=2.7em, text centered, rounded corners, minimum height=1em]
\tikzstyle{line} = [draw, -latex']

\newtheorem{lemma}{Lemma}

\newtheorem{theorem}[lemma]{Theorem}
\newtheorem{corollary}[lemma]{Corollary}
\newtheorem{proposition}[lemma]{Proposition}
\newtheorem{remark}[lemma]{Remark}
\newtheorem{example}[lemma]{Example}
\newtheorem{definition}[lemma]{Definition}

\usepackage{longtable}
\usepackage{listings}

\begin{document}
%
\title{Ordered Orthogonal Array Construction Using LFSR Sequences}
%
%
%

\author{Andr\'e~Guerino~Castoldi, Lucia~Moura,
Daniel~Panario, 
and~Brett~Stevens
\thanks{A. G. Castoldi is with
              Departamento Acad\^emico de Matem\'atica (DAMAT),
              Universidade Tecnol\'ogica Federal do Paraná,
              Via do Conhecimento, Pato Branco, PR, 85503390,
               Brazil.
              email: guerinocastoldi@yahoo.com.br, andrecastoldi@utfpr.edu.br
             }
\thanks{L. Moura is with
           School of Electrical Engineering and Computer Science,
           University of Ottawa, 800 King Edward Ave, Ottawa, ON K1K 6N5, Canada.
             email: lucia@eecs.uottawa.ca}
\thanks{D. Panario and B. Stevens are with
           School of Mathematics and Statistics,
           Carleton University, 1125 Colonel By Drive, Ottawa,
           ON K1S 5B6, Canada.
           emails: daniel@math.carleton.ca, brett@math.carleton.ca}
\thanks{A.~G.~Castoldi was supported by CAPES of Brazil, Science without Borders Program, process 99999.003758/2014-01;
the other three authors by NSERC of Canada.}
\thanks{The work was done while A. G. Castoldi visited the University
of Ottawa as part of his Ph.D.~studies.}
}

\markboth{Last Revised: December, 2016}{Castoldi, Moura, Panario and Stevens.}

\maketitle

\begin{abstract}
We present a new construction of  ordered orthogonal arrays (OOA) of strength $t$
with $(q + 1)t$ columns over a finite field $\mathbb{F}_{q}$ using linear
feedback shift register sequences (LFSRs). OOAs are naturally related to
$(t, m, s)$-nets, linear codes, and MDS codes. Our construction selects
suitable columns from the array formed by all subintervals of length
$\frac{q^{t}-1}{q-1}$ of an LFSR sequence generated by  a
primitive polynomial of degree $t$ over $\mathbb{F}_{q}$. We prove properties about the
relative positions of runs in an LFSR which guarantee that the constructed
OOA has strength $t$. The set of parameters of our OOAs are the same as the
ones given by Rosenbloom and Tsfasman (1997) and Skriganov (2002), but the
constructed arrays are different. We experimentally verify that our OOAs
are stronger than the Rosenbloom-Tsfasman-Skriganov OOAs in the sense that
ours are ``closer'' to being a ``full'' orthogonal array. We also discuss
how our OOA construction relates to previous techniques to build OOAs
from a set of linearly independent vectors over $\mathbb{F}_{q}$, as well
as to hypergraph homomorphisms.
\end{abstract}


%

\section{Introduction}

\IEEEPARstart{O}{rdered} orthogonal arrays (OOA) are a generalization of
orthogonal arrays introduced independently by Lawrence \cite{Lawrence}
and Mullen and Schmid \cite{Mullen}.  A survey of constructions of ordered orthogonal arrays is in \cite[Chapter 3]{Tamar};
we also refer to \cite[Section VI.59.3]{Colburn}.

Let $t$, $m$, $s$, $v$, $\lambda$ be positive integers such that
$2\leq t \leq ms$, and let $N=\lambda v^{t}$.
Let $A$ be an $N\times ms$ array over an alphabet $V$ of size $v$. An $N\times t$ subarray of $A$ is
{\em $\lambda$-covered} if it has each $t$-tuple over $V$ as a row
exactly $\lambda$ times. A set of $t$ columns is {\em $\lambda$-covered} if the $N\times t$ array formed by them
is $\lambda$-covered.  If $\lambda = 1$, we simply say an $N\times t$ subarray or a set of $t$ columns
is {\em covered}. Being $\lambda$-covered is often referred to as having the OA property.

Let $A$ be an array with $ms$ columns labeled by $[m]\times [s]=\{(i, j) : 1 \leq  i \leq m, 1\leq j\leq s  \}$.
A subset $L$ of columns
is {\em left-justified} if $(i, j) \in L$ with $j>1$ implies $(i, j-1) \in L$.
An {\em ordered orthogonal array} $OOA_{\lambda}(N;t,m,s,v)$ is an $N \times ms$ array $A$ with columns labeled by
ordered pairs $(i, j)\in [m]\times [s]$ and with elements from an alphabet $V$ of size $v$, with
the property that every left-justified set $L$ of $t$ columns of $A$ is $\lambda$-covered.
The parameter $t$ is known as the {\em strength} of the OOA.
 Since the parameter $N$ is
determined by the other parameters, we sometimes write
$OOA_{\lambda}(t,m,s,v)$. If $\lambda=1$, we just write $OOA(t,m,s,v)$.
When $s=1$, an $OOA_{\lambda}(N;t,m,1,v)$ is the well-known orthogonal
array $OA_{\lambda}(N;t,m,v)$.

\IEEEpubidadjcol

\begin{figure*}[!t]
\normalsize
\centering
 \begin{equation*}
OOA(3,3,2,2)=\left[\begin{array}{cc|cc|cc}
 $(1,1)$ & $(1,2)$ &  $(2,1)$ & $(2,2)$ &  $(3,1)$ & $(3,2)$  \\ \hline
 0 & 0 &  1 & 1 &  1 & 1  \\
 1 & 0 &  1 & 1 &  0 & 0  \\
 1 & 1 &  1 & 0 &  1 & 0  \\
 1 & 1 &  0 & 1 &  0 & 1  \\
 0 & 1 &  1 & 0 &  0 & 1  \\
 1 & 0 &  0 & 0 &  1 & 1  \\
 0 & 1 &  0 & 1 &  1 & 0  \\
 0 & 0 &  0 & 0 &  0 & 0  \\
\end{array}\right]
\end{equation*}
\caption{A binary  ordered orthogonal array of strength 3.}
\label{fig1}
\end{figure*}

For example, Fig.~\ref{fig1} shows a binary  ordered orthogonal array of strength 3.
The columns of the $OOA(3,3,2,2)$ in Fig.~\ref{fig1} are labeled by $[3]\times[2]=$ $\{(1,1),(1,2),(2,1),(2,2),(3,1),(3,2)\}$.
There are seven left-justified sets of size 3, namely
\begin{table}[!h]
\centering
\begin{tabular}{cc}
  $\{(1,1),(1,2),(2,1)\}$ & $\{(1,1),(1,2),(3,1)\}$ \\
  $\{(1,1),(2,1),(2,2)\}$ & $\{(1,1),(2,1),(3,1)\}$ \\
  $\{(1,1),(3,1),(3,2)\}$ & $\{(2,1),(2,2),(3,1)\}$ \\
  $\{(2,1),(3,1),(3,2)\}$. &  \\
\end{tabular}
\end{table} \\
The $8\times 3$ subarray given by each of them is covered.

Ordered orthogonal arrays are related to the Niederreiter-Rosenbloom-Tsfasman metric and $(t,m,s)$-nets in base $b$
(see the definition of $(t,m,s)$-nets in Section \ref{sec6}).
Rosenbloom and Tsfasman \cite{Rosenbloom} introduced a metric on
linear spaces over finite fields and discussed possible applications of this metric
to interference in parallel channels of communication systems. This metric is commonly known as the
Niederreiter-Rosenbloom-Tsfasman (NRT) metric.
Rosenbloom and Tsfasman \cite{Rosenbloom} and Skriganov \cite{Skriganov} constructed a class
of maximum distance separable (MDS) codes over this metric.
For $q$ a prime power and $s\leq t$, they show that there exists an MDS code with respect
to the NRT metric with length $(q+1)s$, dimension $t$, and minimum distance $(q+1)s-t+1$.
This class of MDS codes is known as Reed-Solomon $s$-codes and they
are equivalent to an $OOA(t,q+1,t,q)$. In this paper we provide a new construction of
OOAs with these parameters.

Niederreiter \cite{N} introduced $(t,m,s)$-nets in base $b$  which we define in Section \ref{sec6}; several
applications of  these objects to numerical integration (quasi-Monte Carlo
methods) can be found in \cite{NiederMC}.
Ordered orthogonal arrays are
a combinatorial characterization of $(t,m,s)$-nets.
Lawrence \cite{Lawrence} and Mullen and Schmid \cite{Mullen} show that there exists a $(t,m,s)$-net
in base $b$ if and only if there exists an \linebreak $OOA_{b^{t}}(m-t,s,m-t,b)$.
 An $OOA(t,q+1,t,q)$ corresponds to a $(0,t,q+1)$-net
in base $q$. The OOA constructed in this paper is a linear OOA, that is, the rows of the $OOA(t,q+1,t,q)$
form a subspace of $\mathbb{F}_{q}^{(q+1)t}$. The $(t,m,s)$-nets corresponding
to linear OOAs are known as digital nets, see for instance
\cite[Section $\mathrm{VI.59}$]{Colburn}.

Our OOAs differ from all previous constructions of ordered
orthogonal arrays  since they are derived from linear feedback
shift register (LFSR) sequences. We review
LFSR concepts in Section \ref{sec2}. Munemasa \cite{Munemasa} first
used LFSR sequences to construct classical orthogonal
arrays; see also \cite{Dewar,Panario}. This was later extended
to covering arrays in \cite{Raaphorst,Tzanakis}.
 A survey of finite field constructions of combinatorial arrays is in \cite{DCCsurvey}.

In order to construct OOAs, we need some properties
of maximum period LFSR sequences that to the best of our knowledge are
identified for the first time in this paper;  see Section \ref{sec3}.
Our main result in this section is Theorem \ref{thm1} which relates runs of zeroes in  the sequence to roots
of a certain polynomial over $\mathbb{F}_{q}$. In Section \ref{sec41}, we give further properties of maximum period LFSR sequences.
Theorem \ref{thm1} is crucial in our construction which is given in Section \ref{sec4}.
The main result of this paper is Theorem \ref{thm3} where we construct an $OOA(t,q+1,t,q)$ using LFSR sequences.
In Section \ref{sec5}, we show experimentally that our method to construct OOAs
covers substantially more $t$-sets of columns than both
Rosenbloom and Tsfasman \cite{Rosenbloom} and  Skriganov \cite{Skriganov} methods.
This also shows that our construction is new and intrinsically different from their construction.
In Section~\ref{sec6}, we relate our construction to other combinatorial structures.
In Subsection~\ref{sec61}, we discuss sets of linearly independent vectors over $\mathbb{F}_q$,
and OOA constructions that utilize linear independence and coding theory.
In Subsection~\ref{sec62}, we consider hypergraphs and homomorphisms to construct OOAs,
showing that this is a framework that includes many previous OOA constructions and non-existence proofs.
In this framework, ours is the first construction for general strength $t$ that satisfies a non-triviality condition asked for by Martin \cite{bill}.

\section{Preliminaries on LFSRs}\label{sec2}

\indent We present concepts and results on finite fields
and linear feedback shift register sequences that are needed to
construct ordered orthogonal arrays of strength $t$ in the subsequent sections.

Let $q$ be a prime power, $t\geq 1$, and let $\mathbb{F}_{q^t}$ be the finite
field of $q^t$ elements. Let $\mathbb{F}_{q^t}^{\times}$ be the multiplicative
group of $\mathbb{F}_{q^t}$. If $\alpha \in \mathbb{F}_{q^t}$ generates
$\mathbb{F}_{q^t}^{\times}$, $\alpha$ is a {\em primitive element} of
$\mathbb{F}_{q^t}$. A polynomial $f \in \mathbb{F}_{q}[x]$ of degree $t\geq 1$
is a {\em primitive polynomial over} $\mathbb{F}_{q}$ if $f$ is the minimal
polynomial over $\mathbb{F}_{q}$ of a primitive element of $\mathbb{F}_{q^{t}}$.

A {\em linear feedback shift register sequence}, henceforth LFSR sequence,
with characteristic polynomial
$f(x)=$ \linebreak $c_{0}+c_{1}x+\cdots+c_{t-1}x^{t-1}+x^{t} \in \mathbb{F}_{q}[x]$ and
initial values $T=(b_{0},\ldots,b_{t-1})\in \mathbb{F}_{q}^{t}$ is a sequence
$S(f,T)=(a_{i})_{i\geq 0}$ over $\mathbb{F}_{q}$ defined as
\setlength{\arraycolsep}{0.0em}
\begin{eqnarray}\label{eq0}
a_{i}  =  \left\{ \begin{array}{ll}
b_{i} & \textrm{if $0\leq i < t$},\\
\displaystyle -\sum_{j=0}^{t-1}c_{j}a_{i-t+j} & \textrm{if $t \leq i$}.
\end{array}\right.
\end{eqnarray}
\setlength{\arraycolsep}{5pt}

A sequence $(a_{i})_{i\geq 0}$ is {\em periodic} if there exists an integer
$r>0$ such that $a_{i+r}=a_{i}$ for all $i\geq 0$; the smallest such $r$ is
the {\em least period} (or simply {\em period} of the sequence).
It is well known (see for example \cite[Theorem 8.33]{Lidl})
that if  $f$ is a primitive polynomial of degree $t$ over $\mathbb{F}_{q}$,
then the LFSR sequence generated by $f$ and nonzero initial values
$T=(b_{0},\ldots,b_{t-1})$ has maximum period $q^{t}-1$.
An LFSR sequence with maximum period is an $m$-sequence.

An important tool when working with LFSRs is the trace function from the
extension field $\mathbb{F}_{q^{t}}$ to the field $\mathbb{F}_{q}$. The
trace function is defined by
$$
\begin{array}{cccl}
\mathrm{Tr} \ : & \! \mathbb{F}_{q^{t}} & \! \longrightarrow & \! \mathbb{F}_{q} \\
& \! x & \! \longmapsto & \! x+x^{q}+x^{q^{2}}+\cdots+x^{q^{t-1}}.
\end{array}
$$
This function is $\mathbb{F}_{q}$-linear; see for example \cite[Theorem 2.23]{Lidl}.

The trace function provides a one-to-one correspondence between
initial values $T\in \mathbb{F}_{q}^{t}$ of an LFSR sequence and elements of $\mathbb{F}_{q^{t}}$, as shown next.

\begin{proposition}\cite[Theorem 8.21]{Lidl}\label{prop2}
Let $f$ be a primitive polynomial of degree $t$ over $\mathbb{F}_{q}$  and $\alpha \in\mathbb{F}_{q^{t}}$
a root of $f$. For any initial values $T=(b_{0},\ldots,b_{t-1})$, there exists a unique element $\gamma
\in \mathbb{F}_{q^{t}}$ such that $b_{i}=\mathrm{Tr}(\gamma \alpha^{i})$ for all $0\leq i < t$. Moreover, the LFSR
$(a_{i})_{i\geq 0}$ generated by $f$ and $T$ has the property that for all  $i\geq 0$, $a_{i}=\mathrm{Tr}(\gamma \alpha^{i})$.
\end{proposition}

For a positive integer $l$ and a sequence $S=(a_{i})_{i\geq 0}$,  we define
\[
C_{i}^{l}(S)=(a_{i}, a_{i+1}, \ldots, a_{i+l-1})
\]
to be the {\em subinterval of $S$ of length $l$ beginning
at position $i$}. Let  $\delta$ $\in \mathbb{F}_{q}$.
The subinterval $C_{i}^{l}(S)$ is a {\em run of $\delta$'s of length $l$} if $a_{i+j}=\delta$, $0\leq j<l$,
and $a_{i-1}\neq \delta$, $a_{i+l}\neq \delta$.

In an LFSR sequence of maximum period, properties $(1)-(4)$ below characterize the run property, also
known as  Golomb's second randomness postulate. Property $(5)$ below is
 Golomb's fourth randomness postulate.

\begin{proposition}\cite[Section 5.2]{Golomb}\label{prop3}
Let $f$ be a primitive polynomial of degree $t$ over $\mathbb{F}_{q}$
and nonzero initial values  $T=(b_{0},\ldots,b_{t-1}) \in \mathbb{F}_{q}^{t}$. In a period of the LFSR
sequence $S(f,T)$ the following properties hold.
\begin{enumerate}
\item[(1)] For $1\leq l \leq t-2$, the runs of every element in $\mathbb{F}_{q}$ of
length $l$ occur $(q-1)^{2}q^{t-l-2}$ times.
\item[(2)] The runs of every nonzero element in $\mathbb{F}_{q}$ of length $t-1$
occur $q-2$ times.
\item[(3)] The runs of the zero element in $\mathbb{F}_{q}$ of length $t-1$ occur
$q-1$ times.
\item[(4)] The run of any nonzero element in $\mathbb{F}_{q}$ of length $t$ occurs
once, and there is no run of zeroes of length $t$.
\item[(5)] Each nonzero $t$-tuple in $\mathbb{F}_{q}^{t}$ appear exactly once as consecutive
elements.
\end{enumerate}
\end{proposition}

The behaviour of the positions of zeroes in any two subintervals of
length $k=\frac{q^{t}-1}{q-1}$ beginning in positions that differ by a
multiple of $k$ is presented in the following result.

\begin{proposition}\cite[Corollary 1]{Raaphorst} \label{prop4}
Let $k=\frac{q^{t}-1}{q-1}$. If $f$ is a primitive polynomial of degree $t$ over $\mathbb{F}_{q}$,
then the LFSR sequence generated by $f$ and  $T\in \mathbb{F}_{q}^{t}$,  $T\neq (0, \ldots,0)$,
has the following properties:
\begin{enumerate}
\item[(1)] For any $i\geq 0$, $C_{i}^{k}(S(f,T))$ contains exactly $\frac{q^{t-1}-1}{q-1}$ zeroes.
\item[(2)] For any $i\geq 0$, $j\geq 0$, the positions of zeroes in $C_{i}^{k}(S(f,T))$ and $C_{i+jk}^{k}(S(f,T))$
are identical.
\end{enumerate}
\end{proposition}

\section{New Properties of LFSRs of Maximum Period}\label{sec3}

\indent In this section, we study properties of linear feedback shift register sequences of maximum period.
They are used to investigate the relationship between runs of elements in $\mathbb{F}_{q}$
in the LFSR sequence generated by a primitive polynomial and nonzero initial values.
In Section \ref{sec4}, we apply these results to prove that the arrays constructed  in that section are
ordered orthogonal arrays of strength $t$.

\begin{proposition}\label{prop5}
Let $S(f,T)=(a_{i})_{i\geq0}$ be the LFSR sequence generated by a primitive polynomial $f$ of degree $t$ over $\mathbb{F}_{q}$
and nonzero initial values $T \in \mathbb{F}_{q}^{t}$. Consider $\alpha \in\mathbb{F}_{q^{t}}$ a root of $f$.
For each $\beta \in \mathbb{F}_{q}^{\times}$,  let $k_{\beta}\in \mathbb{Z}_{q^{t}-1}$ such that $\alpha^{k_{\beta}}(\alpha-\beta)=1$.
Then, in a period of $S(f,T)$
\begin{equation*}
a_{i+1}-\beta a_{i}=a_{i-k_{\beta}}
\end{equation*}
for all $i\geq 0$, where the subscripts of $a$'s are taken modulo $q^{t}-1$.
\end{proposition}

\begin{IEEEproof}
By Proposition \ref{prop2} there exists a unique $\gamma \in \mathbb{F}_{q^{t}}$ such
that $a_{i}=\mathrm{Tr}(\gamma \alpha^{i})$ for all $i\geq 0$.
By the hypotheses, $\alpha-\beta=\alpha^{-k_{\beta}}$. Since the trace function is $\mathbb{F}_{q}$-linear
it follows, for all $i\geq 0$, that
\setlength{\arraycolsep}{0.0em}
\begin{eqnarray*}
a_{i+1}-\beta a_{i} &{}={}& \mathrm{Tr}(\gamma \alpha^{i+1})-\beta \mathrm{Tr}(\gamma \alpha^{i}) \\
                    &{}={}& \mathrm{Tr}(\gamma \alpha^{i}(\alpha-\beta)) \\
                    &{}={}& \mathrm{Tr}(\gamma \alpha^{i-k_{\beta}}) \\
                    &{}={}& a_{i-k_{\beta}}.
\end{eqnarray*}
\setlength{\arraycolsep}{5pt}
\end{IEEEproof}

The previous result means that the difference $a_{i+1}-\beta a_{i}$ is
determined by counting back $k_{\beta}$ positions from position $i$ in a period of $S(f,T)$.

\begin{example}
Let $f(x)=1+x+x^{4}$ be a primitive polynomial over $\mathbb{F}_{2}$ and
$\alpha \in \mathbb{F}_{2^{4}}$ be a root of $f$. Consider the LFSR sequence $S(f,0001)=(a_{i})_{i\geq 0}$.
We take $k_{1}=11\in \mathbb{Z}_{15}$, since $\alpha^{11}(\alpha-1)=1$. By Proposition \ref{prop5},
$a_{i+1}-a_{i}=a_{i-11}$
which means that the difference between consecutive elements of $S(f,0001)$ is
determined by counting back 11 positions from position $i$ in a period of $S(f,0001)$ as shown below 
\begin{figure}[!htp]
\centering
\begin{dependency}[theme = simple]
   \begin{deptext}[column sep=0em]
     $\underbrace{{\bf 0} \ 0010011010}_{11}$ \& {\bf 1} \& {\bf 1} \& 11.  \\
   \end{deptext}
   \depedge{3}{2}{0}
\end{dependency}
\end{figure}

\noindent This property allows us to obtain a run of zeroes in $\mathbb{F}_{2}$
of length $l$ from any run of length $l+1$ by counting back 11 positions as
illustrated below
\begin{figure}[h]
\centering
\begin{dependency}[theme = simple]
   \begin{deptext}[column sep=0em]
     1$\underbrace{{\bf 000}10011010}_{11}$ \& {\bf 1} \& {\bf 1} \& {\bf 1} \& {\bf 1} \& 0.  \\
   \end{deptext}
   \depedge{3}{2}{0}
   \depedge{4}{3}{0}
   \depedge{5}{4}{0}
\end{dependency}
\end{figure}

\noindent In the same way a run of zeroes of length $l$ is obtained from a run of length $l+1$, the process can
be reversed and a run of length $l+1$ is reached by counting forward 11 positions
from a run of zeroes of length $l$.
\end{example}

The next proposition shows that this process can be done for any LFSR sequence
generated by a primitive polynomial.

\begin{proposition} \label{corol1}
Let $f$ be a primitive polynomial of degree $t$ over $\mathbb{F}_{q}$  with $\alpha \in\mathbb{F}_{q^{t}}$
a root of $f$. Let $S(f,T)=(a_{i})_{i\geq 0}$ be the LFSR sequence generated
by $f$ and nonzero initial values $T\in \mathbb{F}_{q}^{t}$.
Consider $\beta \in \mathbb{F}_{q}^{\times}$ and $k_{\beta}\in \mathbb{Z}_{q^{t}-1}$ such that
$\alpha^{k_{\beta}}(\alpha-\beta)=1$. For $l\in\{1,\ldots,t \}$ and $\delta \in \mathbb{F}_{q}$,
if $C_{n}^{l}(S(f,T))$ is a run of $\delta$'s of length $l$, then
$C_{n-k_{\beta}}^{l-1}(S(f,T))$ is a run of $\delta(1-\beta)$'s of length $l-1$.
\end{proposition}

\begin{IEEEproof}
Since $C_{n}^{l}(S(f,T))$ is a run of $\delta$'s of length $l$ then $a_{n-1}\neq \delta$
and $a_{n+l}\neq \delta$. Proposition \ref{prop5}
yields $a_{i}=$ \linebreak $\delta(1-\beta)$ for all $i=n-k_{\beta},\ldots,n-k_{\beta}+l-2$. We claim that
$a_{n-k_{\beta}-1}\neq \delta(1-\beta)$ and $a_{n-k_{\beta}+l-1}\neq \delta(1-\beta)$.
Suppose by contradiction that $a_{n-k_{\beta}-1}= \delta(1-\beta)$. By Proposition \ref{prop5},
$a_{n-k_{\beta}-1}= a_{n}-\beta a_{n-1}$. If $a_{n}-\beta a_{n-1}=\delta(1-\beta)$ and $a_{n}=\delta$,
we have $a_{n-1}=\delta$, which is a contradiction. A similar argument shows that $a_{n-k_{\beta}+l-1}\neq \delta(1-\beta)$.
\end{IEEEproof}

In other words, in a period of $S(f,T)$, a run of $\delta's$ of length $l$ is turned into a run of
$\delta(1-\beta)'s$ of length $l-1$ by making multiple scalar differences between consecutive
elements.
Another way of interpreting the above result is: in a period of an LFSR sequence
when we count back $k_{\beta}$ positions from a run of
$\delta's$ of length $l$, we find a run of $\delta(1-\beta)'s$ of length $l-1$.

\begin{proposition}\label{corol2}
Let $f$ be a primitive polynomial of degree $t$ over $\mathbb{F}_{q}$ and $\alpha \in\mathbb{F}_{q^{t}}$
a root of $f$. Let $S(f,T)=(a_{i})_{i\geq 0}$ be the LFSR sequence generated by $f$ and $T\in \mathbb{F}_{q}^{t}$, $T\neq (0,\ldots, 0)$.
For each $\beta \in \mathbb{F}_{q}^{\times}$, let $k_{\beta}\in \mathbb{Z}_{q^{t}-1}$ such that
$\alpha^{k_{\beta}}(\alpha-\beta)=1$.
If $C_{n-k_{\beta}}^{l-1}(S(f,T))$ is a run of zeroes of length $l-1$, then
$C_{n}^{l}(S(f,T))=(a_{n},\beta a_{n},\beta^2 a_{n},\ldots, \beta^{l-1} a_{n})$. Moreover,
\begin{enumerate}
\item[(1)] if $l=t$, then  $a_{n}\neq 0$;
\item[(2)] if $l\in\{2,\ldots,t\}$ and $a_{n}$ is nonzero (zero) then  $a_{n+1},\ldots, a_{n+l-1}$ are nonzero (zero);
\item[(3)] if $\beta=1$, then $C_{n}^{l}(S(f,T))$ is a run of $a_{n}$'s of length $l$.
\end{enumerate}
\end{proposition}

\begin{IEEEproof}
 Since $C_{n-k_{\beta}}^{l-1}(S(f,T))$ is a run of zeroes of length $l-1$, by Proposition \ref{prop5}
we have that $a_{n+i}=\beta a_{n+i-1}$ for all $i=1,\ldots,l-1$. Therefore,
$a_{n+i}=\beta^{i} a_{n}$ for all $i=1,\ldots,l-1$ and $C_{n}^{l}(S(f,T))=(a_{n},\beta a_{n},\beta^2 a_{n},\ldots, \beta^{l-1} a_{n})$.

$(1)$ Suppose by contradiction that $a_{n}=0$. The subinterval $C_{n}^{t}(S(f,T))$
is a run of zeroes of length $t$, which is a contradiction as such run does not exist by item $(4)$ of Proposition \ref{prop3}.

$(2)$ If $a_{n}\neq 0$ ($a_{n}=0$) it follows that $a_{n+i}\neq 0$ ($a_{n+i}=0$) for all $i=1,\ldots,l-1$.

$(3)$ If $\beta=1$, then $a_{n+i}=a_{n}$ for all $i=1,\ldots,l-1$.
We claim that $a_{n-1}\neq a_{n}$ and $a_{n+l}\neq a_{n}$. Suppose by contradiction that
$a_{n-1}=a_{n}$. Proposition \ref{prop5} implies that $a_{n-k_{1}-1}=a_{n}-a_{n-1}=0$, which
contradicts the hypotheses that $C_{n-k_{\beta}}^{l-1}(S(f,T))$ is a run of zeroes of length $l-1$.
Analogously one can prove that $a_{n+l}\neq a_{n}$.
\end{IEEEproof}

\begin{remark}
Propositions \ref{corol1} and \ref{corol2} hold for a period of an LFSR sequence.
However, $k_{\beta}\in \mathbb{Z}_{q^{t}-1}$ satisfying \linebreak $\alpha^{k_{\beta}}(\alpha-\beta)=1$
can be considered modulo $k=\frac{q^t-1}{q-1}$ and Propositions \ref{corol1} and \ref{corol2}
still hold in a subinterval of $S(f,T)$ of length $k$.
This is possible because of the constant position of the zeroes in subintervals of $S(f,T)$ of length $k$,
given in Proposition \ref{prop4}.
\end{remark}

Let $f(x)=c_{0}+c_{1}x+\cdots+c_{t-1}x^{t-1}+x^{t}$
be a primitive polynomial of degree $t$ over $\mathbb{F}_{q}$  and $\alpha \in \mathbb{F}_{q^{t}}$
be a root of $f$. Let $S(f,T)=(a_{i})_{i\geq 0}$ be the LFSR sequence generated by
$f$ and $T=(b_{0},\ldots,b_{t-1})\in \mathbb{F}_{q}^{t}$, $T\neq (0,\ldots, 0)$.
 For each $\beta \in \mathbb{F}_{q}^{\times}$, let $k_{\beta}\in \mathbb{Z}_{q^{t}-1}$ such that $\alpha^{k_{\beta}}(\alpha -\beta)=1$.
Consider $C_{n}^{l}(S(f,T))$ a run of zeroes of length $l$.  We are interested in the number $z=z(C_{n}^{l},\beta) \in \mathbb{Z}$
such that, for $j=1,\ldots, z$, $a_{n+jk_{\beta}}=0$ and $a_{n+(z+1)k_{\beta}}\neq 0$.
By Proposition \ref{corol2}, we have, for $j=1,\ldots, z$,
$C_{n+jk_{\beta}}^{l+j}(S(f,T))$ is a run of zeroes of length $l+j$,
and $C_{n+(z +1)k_{\beta}}^{l+z +1}(S(f,T))$ is not a run of zeroes.
We show in Theorem \ref{thm1} that $z(C_{n}^{l},\beta)$ is the multiplicity of $\beta$
as a root of a suitable polynomial of degree $t-l-1$ over $\mathbb{F}_{q}$.

Given $C_{n}^{l}(S(f,T))$, a run of zeroes of length $l \in \{0,\ldots,t-3\}$, we now establish
a criterion to know whether $C_{n+k_{\beta}}^{l+1}(S(f,T))$ is a run of zeroes of length $l+1$.
Consider the following subintervals of $S(f,T)$:
\begin{eqnarray*}
& & a_{n-1}\underbrace{0\ldots 0}_{l} a_{n+l} \ldots a_{n+t-1} \\
& \xrightarrow{\text{$k_{\beta}$}} & a_{n+k_{\beta}-1}  \overbrace{a_{n+k_{\beta}} \ldots a_{n+k_{\beta}+t-1}}^{t}a_{n+k_{\beta}+t}.
\end{eqnarray*}
By Eq.~(\ref{eq0}), $a_{n+k_{\beta}+t}$ is written as
\[
a_{n+k_{\beta}+t}=-\sum_{j=0}^{t-1} c_{j}a_{n+k_{\beta}+j},
\]
and therefore
\begin{equation}\label{eq2}
\sum_{j=0}^{t-1} c_{j}a_{n+k_{\beta}+j}+a_{n+k_{\beta}+t}=0.
\end{equation}

The elements $a_{n+k_{\beta}+j}$ of the sequence $S(f,T)$ are expressed in terms of
$\beta$, $a_{n+k_{\beta}}$, and the elements
$a_{n-1+j}$ for  $1\leq j\leq t$ as stated below.

\begin{lemma}\label{lemma1}
Let $f$ be a primitive polynomial of degree $t$ over $\mathbb{F}_{q}$ and  $\alpha\in \mathbb{F}_{q^{t}}$ be
a root of $f$. Let $S(f,T)=(a_{i})_{i\geq 0}$ be the LFSR sequence generated by
$f$ and $T \in \mathbb{F}_{q}^{t}$, $T\neq (0,\ldots,0)$.
 For each $\beta\in \mathbb{F}_{q}^{\times}$, let $k_{\beta}\in \mathbb{Z}_{q^{t}-1}$ satisfying $\alpha^{k_{\beta}}(\alpha-\beta)=1$.
Let $l\in \{0,\ldots,t-3\}$ and $C_{n}^{l}(S(f,T))$ a run of zeroes of length $l$.
 Then
\setlength{\arraycolsep}{0.0em}
\[
a_{n+k_{\beta}+j}{}={}\left\{ \begin{array}{ll}
\beta^{j}a_{n+k_{\beta}} & \textrm{if $j=1,\ldots,l$},\\
\displaystyle \sum_{i=l}^{j-1}\beta^{j-i-1}a_{n+i}+ \beta^{j}a_{n+k_{\beta}} & \textrm{if $j=l+1,\ldots,t$}.
\end{array}\right.
\]
\setlength{\arraycolsep}{5pt}
\end{lemma}

\begin{IEEEproof}
We show first that
\[
a_{n+k_{\beta}+j}=\sum_{i=0}^{j-1}\beta^{j-i-1}a_{n+i}+ \beta^{j}a_{n+k_{\beta}}.
\]
Proposition \ref{prop5} implies that $a_{n+k_{\beta}+1}-\beta a_{n+k_{\beta}}=a_{n}$.
Then $a_{n+k_{\beta}+1}=a_{n}+\beta a_{n+k_{\beta}}$, and the result holds for $j=1$.
For the induction step, suppose that
\[
a_{n+k_{\beta}+j-1}=\sum_{i=0}^{j-2}\beta^{j-i-2}a_{n+i}+\beta^{j-1}a_{n+k_{\beta}}.
\]
Applying Proposition \ref{prop5} again, we can write $a_{n+k_{\beta}+j}=\beta a_{n+k_{\beta}+j-1}+a_{n+j-1}$.
Therefore
\setlength{\arraycolsep}{0.0em}
\begin{eqnarray*}
a_{n+k_{\beta}+j} &{}={}&  a_{n+j-1}+\beta \left( \sum_{i=0}^{j-2}\beta^{j-i-2}a_{n+i}+\beta^{j-1}a_{n+k_{\beta}}\right) \\
                  &{}={}& \sum_{i=0}^{j-1}\beta^{j-i-1}a_{n+i}+ \beta^{j}a_{n+k_{\beta}}.
\end{eqnarray*}
\setlength{\arraycolsep}{5pt}
Since $a_{n+i}=0$ for all $i=0,\ldots,l-1$ the result follows.
\end{IEEEproof}

 Combining Eq.~(\ref{eq2}) and Lemma \ref{lemma1}, the following equation arises
\begin{eqnarray}\label{eq3}
&& \hspace{-0.5cm} \sum_{j=0}^{l}c_{j}\beta^{j}a_{n+k_{\beta}}+\sum_{j=l+1}^{t-1}c_{j}\left( \sum_{i=l}^{j-1}\beta^{j-i-1}a_{n+i}+\beta^{j}a_{n+k_{\beta}}\right)\nonumber \\
&&\hspace{-0.5cm}{+}\:\sum_{i=l}^{t-1}\beta^{t-i-1}a_{n+i}+\beta^{t}a_{n+k_{\beta}}=0.
\end{eqnarray}
Rearranging the coefficients of $a_{n+k_{\beta}}$ in Eq.~(\ref{eq3}), we get
\begin{eqnarray}\label{eq1}
& & a_{n+k_{\beta}}\left(\sum_{j=0}^{t-1}c_{j}\beta^{j}+\beta^{t}\right)+\sum_{j=l+1}^{t-1}c_{j} \sum_{i=l}^{j-1}\beta^{j-i-1}a_{n+i} \nonumber\\
& &{+}\:\sum_{i=l}^{t-1}\beta^{t-i-1}a_{n+i}=0.
\end{eqnarray}
Let $c_{t}=1$. Eq.~(\ref{eq1}) can be rewritten as
\begin{eqnarray*}
& & a_{n+k_{\beta}}f(\beta)+\sum_{j=l+1}^{t-1}c_{j} \sum_{i=l}^{j-1}\beta^{j-i-1}a_{n+i} \\
& &{+}\: c_{t}\sum_{i=l}^{t-1}\beta^{t-i-1}a_{n+i}=0
\end{eqnarray*}
and, therefore,
\begin{equation*}
a_{n+k_{\beta}}f(\beta)+\sum_{j=l+1}^{t}c_{j} \sum_{i=l}^{j-1}\beta^{j-i-1}a_{n+i}=0.
\end{equation*}
Let $P$ be the following polynomial of degree $(t-l-1)$ over $\mathbb{F}_{q}$
\setlength{\arraycolsep}{0.0em}
\begin{eqnarray*}
P(x)&{}={}& \sum_{j=l+1}^{t}c_{j} \sum_{i=l}^{j-1}a_{n+i} x^{j-i-1}\\
    &{}={}& \sum_{j=0}^{t-l-1}c_{j+l+1} \sum_{i=0}^{j}a_{n+l+j-i} x^{i}.
\end{eqnarray*}
\setlength{\arraycolsep}{5pt}
Therefore, $\beta$ is a root of the polynomial $a_{n+k_{\beta}}f(x)+P(x)$ over $\mathbb{F}_{q}$.

Consider $C_{n}^{l}(S(f,T))$ a run of zeroes of length $l$. We recall that
$z=z(C_{n}^{l},\beta) \in \mathbb{Z}$ is defined as the positive integer  such that, for $j=1,\ldots, z$, $a_{n+jk_{\beta}}=0$ and $a_{n+(z+1)k_{\beta}}\neq 0$.

\begin{theorem}\label{thm1}
Let $f$ be a primitive polynomial of degree $t$ over $\mathbb{F}_{q}$ and  $\alpha\in \mathbb{F}_{q^{t}}$ be
a root of $f$. Let $S(f,T)=(a_{i})_{i\geq 0}$ be the LFSR sequence generated by
$f$ and $T \in \mathbb{F}_{q}^{t}$, $T\neq (0,\ldots,0)$.
 For each $\beta\in \mathbb{F}_{q}^{\times}$, let  $k_{\beta}\in \mathbb{Z}_{q^{t}-1}$ satisfying $\alpha^{k_{\beta}}(\alpha-\beta)=1$.
Let $l\in \{0,\ldots,t-3\}$ and $C_{n}^{l}(S(f,T))$ a run of zeroes of length $l$.
\begin{enumerate}
\item[(1)] The subinterval $C_{n+k_{\beta}}^{l+1}(S(f,T))$ is a run of zeroes of length $l+1$
if and only if $\beta$ is a root of the polynomial $P \in \mathbb{F}_{q}[x]$ of degree $(t-l-1)$ given by
\begin{equation}\label{eq4}
 P(x)= \sum_{j=0}^{t-l-1}c_{j+l+1} \sum_{i=0}^{j}a_{n+l+j-i} x^{i}.
\end{equation}
\item[(2)] The number $z(C_{n}^{l},\beta)$ is equal to the multiplicity of $\beta$ as a root of $P(x)$.
\end{enumerate}
\end{theorem}

\begin{IEEEproof}
$(1)$ By the previous arguments, we know that $\beta$ is a root of the polynomial $a_{n+k_{\beta}}f(x)+P(x)$.
If $C_{n+k_{\beta}}^{l+1}(S(f,T))$ is a run of zeroes of length $l+1$, then $a_{n+k_{\beta}}=0$ and
therefore $\beta$ is a root of $P(x)$. Conversely, if $\beta$ is a root of $P(x)$, then $a_{n+k_{\beta}}f(\beta)=0$.
Since $\beta \in \mathbb{F}_{q}^{\times}$ and $f$ is a primitive polynomial over $\mathbb{F}_{q}$, we conclude
that $f(\beta)\neq 0$. Thus $a_{n+k_{\beta}}=0$, and by Proposition \ref{corol2} item $(2)$, it follows that
$C_{n+k_{\beta}}^{l+1}(S(f,T))$ is a run of zeroes of length $l+1$.

$(2)$ Suppose that $C_{n+k_{\beta}}^{l+1}(S(f,T))$ is a run of zeroes of length $l+1$. By item $(1)$,
we have that $\beta$ is a root of the polynomial $P(x)$ given in Eq.~(\ref{eq4}). Now, suppose that
$C_{n+2k_{\beta}}^{l+2}(S(f,T))$ is a run of zeroes of length $l+2$. Applying the previous
arguments for $C_{n+k_{\beta}}^{l+1}(S(f,T))$ and $C_{n+2k_{\beta}}^{l+2}(S(f,T))$,
and item $(1)$, we obtain that $\beta$ is a root of the polynomial
\begin{equation}\label{eq7}
 P_{\beta}(x) =  \sum_{j=0}^{t-l-2}c_{j+l+2} \sum_{i=0}^{j}a_{n+k_{\beta}+l+1+j-i} x^{i} .
\end{equation}
 We claim that $P(x)=(x-\beta)P_{\beta}(x)$. Indeed, we have that $(x-\beta)P_{\beta}(x)$ is equal to
\setlength{\arraycolsep}{0.0em}
\begin{eqnarray*}
\sum_{j=0}^{t-l-2}c_{j+l+2} \sum_{i=0}^{j}a_{n+k_{\beta}+l+1+j-i} (x^{i+1}-\beta x^{i}).
\end{eqnarray*}
\setlength{\arraycolsep}{5pt}
Since $a_{n+k_{\beta}+l}=0$, Proposition \ref{prop5} implies $a_{n+k_{\beta}+l+1}=a_{n+l}$. Applying
Proposition \ref{prop5} several times, we obtain

\setlength{\arraycolsep}{0.0em}
\begin{eqnarray*}
  && \sum_{i=0}^{j}a_{n+k_{\beta}+l+1+j-i} (x^{i+1}-\beta x^{i}) \\
&{}={}&\:{-}a_{n+k_{\beta}+l+1+j} \beta + \sum_{i=1}^{j+1}a_{n+l+j+1-i}x^{i}.
\end{eqnarray*}
\setlength{\arraycolsep}{5pt}
Now, we get
\setlength{\arraycolsep}{0.0em}
\begin{eqnarray}\label{eq6}
(x-\beta)P_{\beta}(x)&{}={}&{-}\:\beta \sum_{j=0}^{t-l-2}c_{j+l+2} a_{n+k_{\beta}+l+1+j}\nonumber\\
& &{+}\:\sum_{j=0}^{t-l-2}c_{j+l+2} \sum_{i=1}^{j+1}a_{n+l+j+1-i}x^{i} .
\end{eqnarray}
\setlength{\arraycolsep}{5pt}
Since $C_{n+k_{\beta}}^{l+1}(S(f,T))$ is a run of zeroes of length $l+1$, Eq.~(\ref{eq0}) implies
\[
a_{n+k_{\beta}+t-1}=-c_{0}a_{n+k_{\beta}-1}-\sum_{j=0}^{t-l-3}c_{j+l+2}a_{n+k_{\beta}+l+1+j}.
\]
Since $a_{n+k_{\beta}}-\beta a_{n+k_{\beta}-1}=a_{n-1}$ and $a_{n+k_{\beta}}=0$,
\setlength{\arraycolsep}{0.0em}
\begin{eqnarray*}
  &&-\beta \sum_{j=0}^{t-l-2}c_{j+l+2} a_{n+k_{\beta}+l+1+j} \\
&{}={}& \beta \Big(c_{0}a_{n+k_{\beta}-1}-c_{0}a_{n+k_{\beta}-1}\\
&&\:{-}\sum_{j=0}^{t-l-3}c_{j+l+2}a_{n+k_{\beta}+l+1+j}-a_{n+k_{\beta}+t-1}\Big)\\
&{}={}& c_{0}\beta a_{n+k_{\beta}-1} =-c_{0}a_{n-1}.
\end{eqnarray*}
\setlength{\arraycolsep}{5pt}
Eq.~(\ref{eq0}) and the fact that
$C_{n}^{l}(S(f,T))$ is a run of zeroes of length $l$ yield
\setlength{\arraycolsep}{0.0em}
\begin{eqnarray*}
-c_{0}a_{n-1}&{}={}&\sum_{j=l+1}^{t-1}c_{j}a_{n+j-1}+a_{n+t-1} \\
             &{}={}&\sum_{j=0}^{t-l-1}c_{j+l+1}a_{n+l+j}.
\end{eqnarray*}
\setlength{\arraycolsep}{5pt}
Eq.~(\ref{eq6}) is written as
\setlength{\arraycolsep}{0.0em}
\begin{eqnarray*}
  && (x-\beta)P_{\beta}(x) \\
&{}={}& \sum_{j=0}^{t-l-1}c_{j+l+1}a_{n+l+j}
  + \sum_{j=0}^{t-l-2}c_{j+l+2} \sum_{i=1}^{j+1}a_{n+l+j+1-i}x^{i} \\
&{}={}& \sum_{j=0}^{t-l-1}c_{j+l+1}a_{n+l+j}
  + \sum_{j=1}^{t-l-1}c_{j+l+1} \sum_{i=1}^{j}a_{n+l+j-i}x^{i}\\
&{}={}& \sum_{j=0}^{t-l-1}c_{j+l+1} \sum_{i=0}^{j}a_{n+l+j-i} x^{i}.
\end{eqnarray*}
\setlength{\arraycolsep}{5pt}
Therefore, $P(x)=(x-\beta)P_{\beta}(x)$. By item $(1)$, we conclude that this process can be repeated as long as $\beta$ is a root of $P(x)$.
Hence, the number $z(C_{n}^{l},\beta)$ is equal to the multiplicity of $\beta$ as root of $P(x)$.
\end{IEEEproof}

\begin{example}
Let $f(x)=2 + 2x + x^{4}$ be a primitive polynomial over $\mathbb{F}_{3}$ and
$\alpha \in \mathbb{F}_{3^{4}}$ a root of $f$. The parameters $k_{1}=27$ and
$k_{2}=76$ satisfy $\alpha^{k_{1}}(\alpha-1)=1$ and $\alpha^{k_{2}}(\alpha-2)=1$,
respectively. A period of $S(f,1000)$ is
\[
1000100110121100210201221010111122201121
\]
\[
2000200220212200120102112020222211102212.
\]
Consider the run $C_{18}^{1}(S(f,1000))$. Since $c_{0}=c_{1}=2$, $c_{2}=c_{3}=0$ and $c_{4}=1$,
the polynomial described by Eq.~(\ref{eq4}) is $P(x)=1+2x^{2}=2(x-1)(x-2)$.
 The polynomial $P_{\beta}$ given in Eq.~(\ref{eq7}) is $P_{1}(x)=2(x-2)$ for $\beta=1$ and $P_{2}(x)=2(x-1)$
for $\beta=2$. Then $P(x)=(x-\beta)P_{\beta}(x)$ for both $\beta=1$ and $\beta=2$.
By Theorem \ref{thm1},
$C_{18+27}^{2}(S(f,1000))$ and $C_{18+76}^{2}(S(f,1000))$ are runs of zeroes of length 2 as illustrated in Fig.~\ref{fig2}.
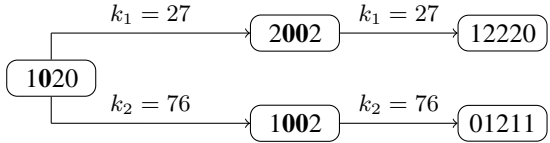
\begin{figure}[h]
    \normalsize
    \centering
\begin{tikzpicture}[scale=1.2, auto]
    \draw
    (0,0) node[block] (run1) {1{\bf 0}20}
    (2.7,0.5) node[block] (run11) {2{\bf 00}2}
    (5,0.5) node[block] (run12) {12220}
    (2.7,-0.5) node[block] (run21) {1{\bf 00}2}
    (5,-0.5) node[block] (run22) {01211};
    \draw[solid] (run1) -- (0,0.5);
    \draw[solid] (run1) -- (0,-0.5);
    \draw[->] (0,0.5) -- node  {\small{$k_{1}=27$}} (run11);
    \draw[->] (0,-0.5) -- node {\small{$k_{2}=76$}} (run21);
    \draw[->] (run11) -- node  {\small{$k_{1}=27$}} (run12);
    \draw[->] (run21) -- node  {\small{$k_{2}=76$}} (run22);
\end{tikzpicture}
\caption{Runs of zeroes obtained from the run $C_{18}^{1}(S(f,1000))$.}
\label{fig2}
\end{figure}

Now, choose the run $C_{27}^{1}(S(f,1000))$.  The polynomial $P$ in this case is $P(x)=1+x+x^{2}=(x-1)^{2}$.
The polynomial $P_{\beta}$ given in Eq.~(\ref{eq7}) is $P_{1}(x)=x-1$ for $\beta=1$ and $P_{2}(x)=x$
for $\beta=2$. Thus, $P(x)=(x-1)P_{1}(x)$ and $P(x)\neq (x-2)P_{2}(x)$.
By Theorem \ref{thm1}, $C_{27+27}^{2}(S(f,1000))$ is a run of zeroes of length 2 and $C_{27+76}^{2}(S(f,1000))$
is not a run of zeroes. Moreover, $C_{27+2\cdot 27}^{3}(S(f,100))$
is a run of zeroes of length 3. Fig.~\ref{fig3} shows the runs of zeroes obtained from the run $C_{27}^{1}(S(f,1000))$.
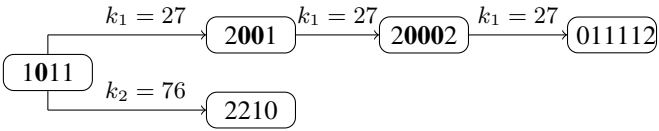
\begin{figure}[!h]
    \normalsize
    \centering
\begin{tikzpicture}[scale=1, auto]
    \draw
    (0,0) node[block] (run1) {1{\bf 0}11}
    (2.7,0.5) node[block] (run11) {2{\bf 00}1}
    (5,0.5) node[block] (run12) {2{\bf 000}2}
    (7.5,0.5) node[block] (run13) {011112}
    (2.7,-0.5) node[block] (run21) {2210};
    \draw[solid] (run1) -- (0,0.5);
    \draw[solid] (run1) -- (0,-0.5);
    \draw[->] (0,0.5) -- node  {\hspace{0.5cm}\small{$k_{1}=27$}} (run11);
    \draw[->] (0,-0.5) -- node {\hspace{0.5cm}\small{$k_{2}=76$}} (run21);
    \draw[->] (run11) -- node  {\small{$k_{1}=27$}} (run12);
    \draw[->] (run12) -- node  {\small{$k_{1}=27$}} (run13);
\end{tikzpicture}
\caption{Runs of zeroes obtained from the run $C_{27}^{1}(S(f,1000))$.}
\label{fig3}
\end{figure}
\end{example}

\section{Further Results of LFSR Sequences}\label{sec41}

 In this section, we show that there exists a
bijection between the runs of nonzero elements of length greater than $l$ and
the runs of zeroes of length exactly $l$. Although we do not use this result in this paper, we think
it may be of independent interest as a new combinatorial property of LFSRs.

Proposition \ref{prop3} states, for $l\in \{1,\ldots,t-2\}$,
there exists exactly $(q-1)^{2}q^{t-l-2}$ runs of zeroes of length $l$. Let $R(l)$ be the total number
of runs of nonzero elements of $\mathbb{F}_{q}$ of length greater than $l$. Proposition \ref{prop3} implies
\setlength{\arraycolsep}{0.0em}
\begin{eqnarray*}
R(l) &{}={}& (q-1)\sum_{i=l+1}^{t-2}(q-1)^{2}q^{t-i-2}+(q-1)^2 \\
     &{}={}& (q-1)^{2}\Big(1+(q-1)\sum_{i=l+1}^{t-2}q^{t-i-2}\Big)  \\
     &{}={}& (q-1)^{2}q^{t-l-2}.
\end{eqnarray*}
\setlength{\arraycolsep}{5pt}
Thus $R(l)$ is equal to the number of runs of zeroes of length $l$.
For $l=t-1$ the same property holds. The application of
Propositions \ref{corol1} and \ref{corol2} and the counting argument on $R(l)$ suggest a bijection
between the runs of nonzero elements of length greater than $l$ and
the runs of zeroes of length exactly $l$. We give this bijection next.

\begin{proposition}\label{prop7}
Let $1\leq l\leq t-1$. In a period of an LFSR sequence $S(f,T)$, there is a bijection between runs of zeroes of length $l$
and runs of nonzero elements of $\mathbb{F}_{q}$ with length larger than $l$.
Moreover, this bijection is such that the difference of indices between the
start of the runs is a multiple of $k_{1}$, where
$k_{1}\in \mathbb{Z}_{q^{t}-1}$ satisfies $\alpha^{k_{1}}(\alpha -1)=1$.
\end{proposition}

\begin{IEEEproof}
Let $C_{n}^{l}(S(f,T))$ be a run of zeroes of length $l$ and $k_{1}\in \mathbb{Z}_{q^{t}-1}$ such that $\alpha^{k_{1}}(\alpha-1)=1$.
By iterating Proposition \ref{corol2}, there exists an integer $j$ such that $C_{n+jk_{1}}^{l+j}(S(f,T))$
is a run of a nonzero element in $\mathbb{F}_{q}$ beginning at position $n+jk_{1}$, and
$C_{n+ik_{1}}^{l+i}(S(f,T))$ is a run of zeroes of length $l+i$ for $i=1,\ldots,j-1$. Furthermore, $j \leq t-l$
and let $l'\in \mathbb{Z}$ such that $j=l'-l$.
Let $S(l)$ be the set formed by the starting positions of the runs of zeroes of length $l$.
Define the map $g(i)=i+jk_{1}=i+(l'-l)k_{1}$ where $i\in S(l)$.

For the inverse map, let $C_{n'}^{l'}(S(f,T))$ be a run of a  nonzero element in $\mathbb{F}_{q}$ of length $l'>l$.
We get $C_{n'-(l'-l)k_{1}}^{l}(S(f,T))$, a run of zeroes of length $l$, by iterating Proposition \ref{corol1} exactly $l'-l$ times.
Let $T(l)$ be the set formed by the starting positions of the runs of a nonzero element in $\mathbb{F}_{q}$ of length $l'>l$.
Consider the map $h(i)=i-(l'-l)k_{1}$ where $i\in T(l)$.

Propositions \ref{corol1} and \ref{corol2} describe inverse processes and so the maps $g$ and $h$ are inverse maps.
\end{IEEEproof}

\begin{example}
Consider the primitive polynomial $f(x)=$ \linebreak $ 1 + 2 x + x^{3}$ over $\mathbb{F}_{3}$.
A period of $S(f,100)$ is given by:
$$10020212210222001012112011.$$
Then $k_{1}=23$ satisfies $\alpha^{23}(\alpha-1)=1$, where $\alpha \in \mathbb{F}_{3^{3}}$
is a root of $f$. In Fig.~\ref{fig0},
the period of the sequence $S(f,100)$ is represented twice and clockwise ordered.
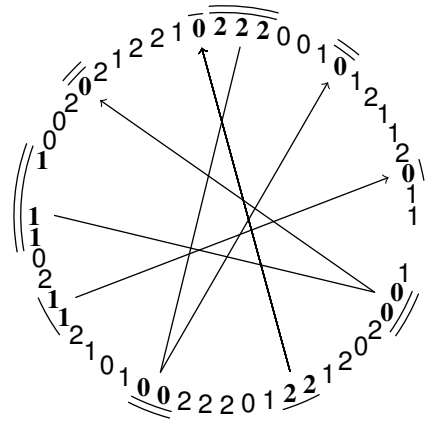
\begin{figure}[h]
\centering
\begin{tikzpicture}[scale=0.9,line cap=round,line width=0.5pt]
\foreach \angle / \label in
{0/1, 6.5/1, 13/{\bf 0}, 19.5/2, 26/1, 32.5/1, 39/2, 45.5/1, 52/{\bf 0},
58.5/1, 65/0, 71.5/0, 78/{\bf 2}, 84.5/{\bf 2}, 91/{\bf 2}, 97.5/{\bf 0}, 104/1, 110.5/2,
117.5/2, 124/1, 130.5/2, 137/{\bf 0}, 143.5/2, 150/0, 156.5/0, 163/{\bf 1},
180/{\bf 1}, 186.5/{\bf 1}, 193/0, 199.5/2, 206/{\bf 1}, 212.5/{\bf 1}, 219/2, 225.5/1, 232/0, 238.5/1,
245/{\bf 0}, 251.5/{\bf 0}, 258/2, 264.5/2, 271/2, 277.5/0, 284/1, 290.5/{\bf 2}, 297/{\bf 2}, 303.5/1,
310/2, 316.5/0, 323/2, 329.5/{\bf 0}, 336/{\bf 0}, 342.5/1}
\draw (\angle:2.8cm) node{\textsf{\label}};
\draw  (160:3cm) arc (160:190:3cm);
\draw  (160:3.1cm) arc (160:190:3.1cm);
\draw  (325:3cm) arc (325:338:3cm);
\draw  (325:3.1cm) arc (325:338:3.1cm);
\draw  (133:3cm) arc (133:140:3cm);
\draw  (133:3.1cm) arc (133:140:3.1cm);
\draw  (287:3cm) arc (287:298:3cm);
\draw  (96:3cm) arc (96:100:3cm);
\draw  (75:3cm) arc (75:94:3cm);
\draw  (75:3.1cm) arc (75:94:3.1cm);
\draw  (243:3cm) arc (243:255:3cm);
\draw  (243:3.1cm) arc (243:255:3.1cm);
\draw  (50:3cm) arc (50:57:3cm);
\draw  (50:3.1cm) arc (50:57:3.1cm);
\draw  (204:3cm) arc (204:216:3cm);
\draw  (10:3cm) arc (10:16:3cm);
\foreach \angle in {13,52,84.5,97.5,180,137,209,248,293,333,175}
\draw [<-] (97.5:2.5cm) -- (293:2.5cm);
\draw [<-] (137:2.5cm) -- (333:2.5cm);
\draw  (333:2.5cm) -- (180:2.5cm);
\draw [<-] (13:2.5cm) -- (209:2.5cm);
\draw [<-] (52:2.5cm) -- (248:2.5cm);
\draw  (248:2.5cm) -- (84.5:2.5cm);
\end{tikzpicture}
\caption{A one-to-one correspondence between runs of zeroes of length $1$ and
runs of nonzero elements of $\mathbb{F}_{3}$ of length larger than $1$.}
\label{fig0}
\end{figure}
The runs highlighted by two arcs represent how the runs of zeroes of length 1 are obtained by counting counterclockwise 23 positions twice
from a run of a nonzero element of length 3. The runs highlighted by one arc represent how the runs of zeroes of length 1
are obtained by counting counterclockwise 23 positions from a run of a nonzero element of length 2.
This illustrates the one-to-one correspondence between runs of zeroes of length $1$ and
runs of nonzero elements of $\mathbb{F}_{3}$ of length larger than $1$, established in Proposition \ref{prop7}.
\end{example}

\section{Ordered Orthogonal Arrays from LFSRs}\label{sec4}

\indent Linear feedback shift register sequences of maximum period are
used to construct orthogonal arrays \cite{Dewar,Munemasa,Panario} and covering arrays \cite{Raaphorst,RaaphorstT,Tzanakis}.
A subinterval array
of the sequence is the key for building such arrays. We construct an ordered
orthogonal array by choosing suitable columns of this subinterval array.

Let $f$ be a primitive polynomial of degree $t\geq 3$ and $\alpha\in \mathbb{F}_{q^{t}}$ a
root of $f$. Let $T \in \mathbb{F}_{q}^{t}$ be nonzero initial values for the sequence $S(f,T)$ generated
by $f$. Let $k=\frac{q^{t}-1}{q-1}$ and consider the following $q^{t}\times k$ array
\[
M=M(f,T)=\left[\begin{array}{c}
  C_{0}^{k}(S(f,T)) \\
  C_{1}^{k}(S(f,T)) \\
  \vdots  \\
  C_{q^{t}-2}^{k}(S(f,T)) \\
  0, 0, \cdots, 0
\end{array}
\right],
\]
where $C_{i}^{k}(S)$ is the subinterval of $S$ of length $k$ beginning
at position $i$. Matrix $M$ is the {\em subinterval array of $f$}.
Label the columns of $M$ by $\mathbb{Z}_{k}$. A characterization of which sets of
$t$ columns of $M$ are covered is given in \cite{Raaphorst}.

\begin{theorem}\cite[Theorem 2]{Raaphorst}\label{thm2}
Let $f$ be a primitive polynomial of degree $t\geq 3$ over $\mathbb{F}_{q}$  and $\alpha \in\mathbb{F}_{q^{t}}$
a root of $f$. Let $k=\frac{q^{t}-1}{q-1}$, and let $M$ be the $q^t\times k$ subinterval array of $f$.
The following are equivalent:
\begin{enumerate}
\item[(1)] A set of $t$ columns $\{i_{1},\ldots,i_{t}\}$ is covered in $M$.
\item[(2)] There is no row $r$ other than the all-zero row of $M$ such that $r_{i_{1}}=\cdots=r_{i_{t}}=0$  .
\item[(3)] The set $\{\alpha^{i_{1}},\ldots,\alpha^{i_{t}}\}$ is linearly independent over $\mathbb{F}_{q}$.
\end{enumerate}
\end{theorem}

\begin{proposition}\label{corol4}
Any subarray of $M$ formed by $t$ consecutive columns of $M$ is covered.
\end{proposition}

\begin{IEEEproof}
It is an immediate consequence of Proposition \ref{prop3}, item $(5)$, and the definition of $M(f,T)$.
\end{IEEEproof}

We are ready to construct ordered orthogonal arrays from the subinterval array of
a primitive polynomial.

\begin{theorem}\label{thm3}
Let $f$ be a primitive polynomial of degree $t\geq 3$
over $\mathbb{F}_{q}$  and $\alpha \in \mathbb{F}_{q^{t}}$ be a root of $f$.
Consider an array $A$ with columns labeled by $[q+1]\times[t]=$ \linebreak $\{(i,j): 1\leq i\leq q+1, 1\leq j\leq t\}$ where
\begin{enumerate}
\item[(1)] columns labeled by $(1,1),\ldots,(1,t)$ correspond to the columns of $M$ labeled by $t-1, \ldots, 0$, respectively;
\item[(2)] columns labeled by $(2,1),\ldots,(2,t)$ correspond to the columns of $M$ labeled by $t,\ldots,2t-1$, respectively; and
\item[(3)] columns labeled by $(i,1),(i,2),\ldots,(i,t)$, for \linebreak $i \in \{3,\ldots, q+1\}$, correspond to the columns of $M$ labeled by
$t+k_{\beta},t+2k_{\beta},\ldots,t+tk_{\beta}$, for each $\beta \in \mathbb{F}_{q}^{\times}$, respectively.
\end{enumerate}
Then, the array $A$ is an $OOA(t,q+1,t,q)$.
\end{theorem}

\begin{IEEEproof}
Let $L$ be a set of $t$ columns in $M$ corresponding to a left-justified set of $t$ columns of $A$.
Given any row $r=(r_{0} \ \ldots \ r_{k-1})$ distinct from the all zero row, we prove that $r_{i}\neq 0$ for some $i\in L$,
and by  Theorem \ref{thm2} the subarray of $M$ labeled by $L$ is covered.
Four cases need to be considered:

{\it Case 1:} $L \subset \{0,\ldots,2t-1\}$. Since $L$ is a left-justified set,
the elements of $L$ are consecutive elements of \linebreak $\{0,\ldots,2t-1\}$.
Then, $L$ is a left-justified set that labels consecutive $t$ columns of $M$, and by Proposition \ref{corol4}
these $t$ columns of $M$ are covered.

{\it Case 2:} $|L \cap \{0,\ldots,2t-1\}|=t-1$. In this case, the left-justified set $L$ has
the form $L=$ \linebreak $\{n,\ldots, n+t-2\}\cup \{t+k_{\beta}\}$ for some $n\in \{1,\ldots,t\}$ and \linebreak $\beta \in
\mathbb{F}_{q}^{\times}$.
If $r_{i}\neq 0$ for some $i\in \{n,\ldots, n+t-2\}$ we have nothing to prove.
Otherwise, $r_{i}=0$ for all $i\in \{n,\ldots, n+t-2\}$. Since an LFSR sequence does not contain a run of
zeroes of length $t$, the consecutive zeroes in the positions $n,\ldots, n+t-2$ form a run of zeroes of length $t-1$.
Proposition \ref{corol2}, items $(1)$ and $(2)$, yield $r_{n+k_{\beta}+j}\neq 0$ for all $j=0,\ldots,t-1$.
In particular, for $j=t-n$ we conclude that $r_{t+k_{\beta}}\neq 0$ as desired.

{\it Case 3:}  $1\leq |L \cap \{0,\ldots,2t-1\}|<t-1$.
We can write $L=L_{1}\cup L_{2}$, where
$L_{1}\subset \{2,\ldots, 2t-3\}$
and \linebreak $L_{2}\subset \{t+jk_{\beta}:  \ \beta \in \mathbb{F}_{q}^{\times}, \ 1\leq j \leq t\}$.
 The set $L_{1}$ is formed by consecutive elements of $\{2,\ldots, 2t-3\}$.
Furthermore, since $L$ is a left-justified set, $t-1$  or $t$ belongs to $L_{1}$. Let $l=|L_{1}|$.
Without loss of generality we assume that the consecutive elements
in $L_{1}$ form a run of zeroes of length $l$
$$ r_{n-1}\underbrace{0\ldots 0}_{l}r_{n+l}r_{n+l+1}\ldots r_{n+t-2}r_{n+t-1},$$
where $n\in \{2,\ldots,t\}$. If $r_{n},\ldots, r_{n+l-1}$ were not all zero, then the proof of this case would be complete.
By Theorem \ref{thm1}, item $(2)$, the number  $z(C_{n}^{l},\beta)$
is equal to the multiplicity of $\beta \in\mathbb{F}_{q}^{\times}$ as a root of the polynomial
\begin{equation*}
 P(x)= \sum_{j=0}^{t-l-1}c_{j+l+1} \sum_{i=0}^{j}r_{n+l+j-i} x^{i},
\end{equation*}
where $c_{0},\ldots, c_{t-1}$ are the coefficients of $f$.
By Proposition \ref{corol2}, items $(1)$ and $(2)$, if $\beta$ is a root of $P(x)$
with multiplicity $z$, then $r_{t+jk_{\beta}}=0$ for $1\leq j \leq z$,
and $r_{t+(z+1)k_{\beta}}\neq 0$.
Since the number of roots of $P(x)$ in $\mathbb{F}_{q}$
counting with multiplicity is at most $t-l-1$, the number of elements $i\in L_{2}$ such that $r_{i}=0$
is at most $t-l-1$. Therefore, the number of zeroes in row $r$ in positions $i\in L$ is at most $t-1$.

{\it Case 4:} $L \cap \{0,\ldots, 2t-1\}=\emptyset$.
We consider two subcases.

If $r_{t}\neq 0$, then we have a run of zeroes of length 0 beginning at position $t$.
By Theorem \ref{thm1}, item $(2)$, the number  $z(C_{t}^{0},\beta)$
is equal to the multiplicity of $\beta \in\mathbb{F}_{q}^{\times}$ as a root of the polynomial
\begin{equation*}
P(x)= \sum_{j=0}^{t-1}c_{j+1} \sum_{i=0}^{j}r_{t+j-i} x^{i}.
\end{equation*}
Since the number of roots of $P(x)$ in $\mathbb{F}_{q}$
counting with multiplicity is at most $t-1$, the number of elements $i\in L$ such that $r_{i}=0$
is at most $t-1$.

If $r_{t}=0$, then there exists $n\in \{2,\ldots,t\}$ such that $C_{n}^{l}=(r_{n},\ldots,r_{t},\ldots,r_{n+l-1})$
is a run of zeroes of length $l\geq 1$. Now, we can apply the same argument as we applied in Case 3 and conclude that
the number of zeroes in row $r$ in positions $i\in L$ is at most $t-1$.
\end{IEEEproof}

The previous theorem constructs an OOA with the same parameter sets
as Skriganov \cite{Skriganov} which generalizes and also covers the range given
by Rosenbloom and Tsfasman \cite{Rosenbloom}.
However, our construction yields different OOAs than the ones in \cite{Skriganov}.
In the next section, we compare characteristics of both constructions,
showing that ours cover a larger number of $t$-sets of columns beyond the
coverage required for the  left-justified sets of the set $[q+1]\times[t]$.
In this sense, the array $A$ in Theorem \ref{thm3} is ``closer'' to being a full orthogonal
array than the one given in \cite{Skriganov}.

\section{A comparative analysis of the OOA constructions}\label{sec5}

In this section, we compare our construction of ordered orthogonal arrays
given in Theorem \ref{thm3}
with the construction of Skriganov \cite{Skriganov}. Since the
construction of  Skriganov \cite{Skriganov}
is broader than Rosenbloom and Tsfasman \cite{Rosenbloom} construction, we
use the former for this comparison. We refer to the construction given in
Theorem \ref{thm3}
as {\em the RUNS construction} and to the one by Rosenbloom and Tsfasman
\cite{Rosenbloom} and
Skriganov \cite{Skriganov} as {\em the RTS construction}.

One criteria we use for comparing different \linebreak $OOA(t,q+1,t,q)$ is their extent of
coverage, that is, the total number of $t$-sets of columns that are covered.
In both constructions, we know that the columns labeled by left-justified sets of
size $t$ are
covered, but there may be many other $t$-sets of columns
that are covered. We were not able to
determine this quantity in general for {\it RUNS} and {\it RTS}, but for some values
of $t$ and $q$, we computed the total number of $t$-sets of columns covered
by {\it RUNS} using  all distinct primitive polynomials and by {\it RTS}, as shown in Table \ref{tab1}.
These experiments  show that {\it RUNS} covers many more $t$-sets of columns than {\it RTS}.

For every $q$ and $t$ considered in Table \ref{tab1} we show statistics on
the ratio
of $t$-sets covered
by RUNS over all possible $t$-sets of columns, for all primitive
polynomials of degree $t$ over $\mathbb{F}_{q}$.
In this table,
$\#f$ is the number of primitive polynomials of degree $t$ over
$\mathbb{F}_{q}$;
{\it RUNS}$_{\min}$, {\it RUNS}$_{\max}$ and {\it RUNS}$_{avg}$ give the minimum, maximum and
average ratio of coverage of {\it RUNS} for
all primitive polynomial for $t$ and $q$;
{\it RTS}  is the ratio of $t$-sets of columns covered by the {\it RTS} construction.

\begin{table*}[!t]
\caption{Ratio of covered $t$-sets for OOAs constructions.}
\label{tab1}
\centering
\begin{tabular}{|ccc|ccc|c|}
\hline\noalign{\smallskip}
$t$ & $q$ & $\#f$  & {\it RUNS}$_{\min}$  & {\it RUNS}$_{\max}$ &  {\it RUNS}$_{avg}$  &
$RTS$  \\
\noalign{\smallskip}\hline\noalign{\smallskip}
3  & 2  & 2  & 0.595238 & 0.595238 &  0.595238  & 0.464286 \\
3  & 3  & 4  & 0.709091 & 0.740909 &  0.723864  & 0.545455 \\
3  & 5  & 20 & 0.810049 & 0.839461 &  0.824387  & 0.573529 \\
3  & 7  & 36 & 0.853261 & 0.889822 &  0.867054  & 0.583004 \\ \hline
4  & 2  & 2  & 0.484848 & 0.523232 &  0.50404    & 0.345455 \\
4  & 3  & 8  & 0.588462 & 0.702747 &  0.632143  & 0.325824 \\
4  & 5  & 48 & 0.776774 & 0.801525 &  0.78791    & 0.449558 \\ \hline
5  & 2  & 6  & 0.38628  & 0.46953  &  0.444388  & 0.196803 \\
5  & 3  & 22 & 0.602941 & 0.660733 &  0.635038  & 0.243292 \\ \hline
6  & 2  & 6  & 0.38914  & 0.446509 &  0.410032  & 0.135693 \\
6  & 3  & 48 & 0.453089 & 0.633845 &  0.59164    & 0.205296 \\ \hline
7  & 2  & 18 & 0.308763 & 0.423719 &  0.363138  & 0.0897059 \\
\noalign{\smallskip}\hline
\end{tabular}
\end{table*}

Table \ref{tab1} shows that in both constructions
many $t$-sets of columns are covered in addition to the left-justified ones.
For all pairs of parameters $(t,q)$ the experiments show that the number
of $t$-sets of columns covered by {\it RUNS} is always greater than this
number for {\it RTS}. For {\it RUNS}, Table \ref{tab1} shows that the
minimum ratio and the maximum ratio of $t$-sets of columns covered are
relatively close, and we conjecture that, for $t$ and $q$ fixed, there
exists a nontrivial lower bound for the number of $t$-sets of columns
covered in all ordered orthogonal arrays obtained with the {\it RUNS}
construction. Finally, when we fix $t$ and vary $q$, the average ratio of
$t$-sets of columns covered by the {\it RUNS} construction grows faster
than the ratio of $t$-sets of columns covered by the {\it RTS} construction.
Moreover, when we fix $q$ and increase $t$, the ratio of $t$-sets of
columns covered by the {\it RTS} construction decreases faster than the
average ratio of $t$-sets of columns covered by the {\it RUNS} construction.

\section{Ordered orthogonal arrays and other combinatorial structures} \label{sec6}

In this section, we relate our methods to other combinatorial structures.  In the first subsection, we look at other constructions that depend on linear independence including coding theory constructions.  In the second subsection, we look at hypergraphs and homomorphisms.

Ordered orthogonal arrays are a combinatorial characterization of
$(t,m,s)$-nets \cite{Lawrence,Mullen}.
Lawrence \cite[Theorem 4.1]{Lawrence} and Mullen
and Schmid \cite[Theorem 7]{Mullen} independently show the equivalence
between $(t,m,s)$-nets and OOAs.
To state this equivalence, we define the concept of $(t, m, s)$-net in base $b$.

Let $[0, 1)^s$ be the half-open unit cube of dimension $s$ and suppose numerical computation
is to be done in base $b \geq 2$. An {\em elementary interval in base b} in $[0, 1)^s$ is an
euclidean set of the form
\[
E = \prod_{i=1}^{s}\left[ \dfrac{a_{i}}{b^{d_{i}}},\dfrac{a_{i+1}}{b^{d_{i}}} \right]
\]
where, for each $i$, $d_{i} \geq 0$ and  $0\leq a_{i} < b^{d_{i}}$.
The volume of $E$ is $b^{-\sum d_{i}}$.
Let $s \geq 1$, $b \geq 2$, and $m \geq t \geq 0$ be integers. A {\em $(t, m, s)$-net in base $b$} is a multiset
$\mathcal{N}$ of $b^{m}$ points in $[0, 1)^s$ with the property that every elementary interval in base $b$
of volume $b^{t-m}$ contains precisely $b^t$ points from $\mathcal{N}$.

\begin{theorem}(\hspace{-0.2mm}\cite{Lawrence,Mullen})
Let $s\geq 1$, $b\geq 2$, $t\geq 0$ and $m$ be integers, and assume
that $m \geq t+1$ to avoid degeneracy. Then there exists a
$(t,m,s)$-net in base $b$ if and only if there exists an \linebreak
$OOA_{b^{t}}(b^{m};m-t,s,m-t,b)$.
\end{theorem}

An ordered orthogonal array $OOA(t,m,s,q)$ over $\mathbb{F}_{q}$
is linear if the rows of the OOA form a subspace of $\mathbb{F}_{q}^{ms}$.
Linear OOAs correspond to digital nets, see for instance
\cite[Section $\mathrm{VI.59}$]{Colburn}. Thus our \linebreak $OOA(t,q+1,t,q)$
is a linear OOA and corresponds to a $(0,t,q+1)$-net in base $q$,
which is consequently a digital net.

\subsection{Ordered orthogonal arrays and sets of independent vectors over $\mathbb{F}_{q}$}\label{sec61}

Ordered orthogonal arrays can be constructed from a set of vectors in $\mathbb{F}_{q}^{t}$ such that
any subset of $t$ vectors is linearly independent over $\mathbb{F}_{q}$. We show, in this subsection, that the RUNS construction
of OOAs uses  a set of vectors in $\mathbb{F}_{q}^{t}$ such that not all choices of $t$ vectors form
a linearly independent set over $\mathbb{F}_{q}$.

A method to construct digital nets in prime power bases is introduced
in \cite{Maha}. This construction makes use of sets of independent
vectors over finite fields. An {\it $(n,t)$-set} in $\mathbb{F}_{q}^{u+t}$
is a set of $n$ vectors in $\mathbb{F}_{q}^{u+t}$ such that any $t$
of them are linearly independent over $\mathbb{F}_{q}$. For our purpose,
we state \cite[Theorem 1]{Maha} in terms of ordered orthogonal arrays.

\begin{theorem}\cite[Theorem 1]{Maha} \label{thm4}
Let $q$ be a prime power, and let $n,u \geq 0$ and $t\geq 2$ be integers.
Given an $(n,t)$-set in $\mathbb{F}_{q}^{u+t}$, a linear
$OOA_{q^{u}}(q^{t+u};t,m,t,q)$ can be constructed over $\mathbb{F}_{q}$ with

\setlength{\arraycolsep}{0.0em}
\begin{eqnarray*}
m  =  \left\{ \begin{array}{ll}
\left\lfloor \dfrac{n-1}{h} \right\rfloor  & \textrm{if $t=2h+1$},
\vspace{0.25cm} \\
\left\lfloor \dfrac{n}{h} \right\rfloor & \textrm{if $t=2h$}.
\end{array}\right.
\end{eqnarray*}
\setlength{\arraycolsep}{5pt}
\end{theorem}

In Corollary~\ref{shortOOA}, we give an
upper bound for the parameter $m$ of the $OOA(t,m,t,q)$ constructed
in Theorem \ref{thm4}, depending on $q$ and $t$.
We first state a classical bound on orthogonal arrays,
before we show the mentioned upper bound.

\begin{theorem}\cite[Theorem 2.19]{Hedayat}\label{thm5}
In an $OA(q^{t};t,n,q)$, the following inequalities hold
\[
\begin{array}{ccll}
n & \leq & t+1  & \textrm{if $q\leq t$}, \\
n & \leq & q+t-2  & \textrm{if $3\leq t <q$ and $q$ is odd}, \\
n & \leq & q+t-1 & \textrm{otherwise}.
\end{array}
\]
\end{theorem}

The concept of $(n,t)$-set in $\mathbb{F}_{q}^{u+t}$ is
connected to the theory of error-correcting codes. The existence
of an $(n,t)$-set in $\mathbb{F}_{q}^{u+t}$ is equivalent to the
existence of a linear \linebreak $[n,n-(u+t),t+1]$-code over $\mathbb{F}_{q}$, see for example \cite[Theorem 5.3.7]{Roman}.
Using this connection of $(n,t)$-sets and coding theory, and Theorem \ref{thm5} we derive
the following upper bound on $n$.

\begin{proposition}\label{boundlinear}
Let $n \geq t \geq 1$, and let $q$ be a prime power.
If there exists an $(n,t)$-set in $\mathbb{F}_q^t$, then there exists an $OA(q^{t};t,n,q)$.
In particular, $n\leq q+t-1$.
\end{proposition}

\begin{IEEEproof}
We first show that the existence of an $(n,t)$-set in $\mathbb{F}_{q}^{t}$
implies the existence of an $OA(q^{t};t,n,q)$.
The existence of an $(n,t)$-set in $\mathbb{F}_{q}^{t}$ is equivalent to the
existence of a linear $[n,n-t,t+1]$-code over $\mathbb{F}_{q}$,
which is consequently an MDS (maximum distance separable) code.
It is known that if a linear code over $\mathbb{F}_{q}$ is an MDS code, then its dual code is also a linear MDS code.
Thus if there exists an MDS $[n,n-t,t+1]$-code over $\mathbb{F}_{q}$, then
there exists an  MDS $[n,t,n-t+1]$-code over $\mathbb{F}_{q}$.
The existence of a linear MDS $[n,t,n-t+1]$-code over $\mathbb{F}_{q}$ implies the existence
of an $OA(q^{t};t,n,q)$, see for instance \cite[Theorem 4.6]{Hedayat}.
Now, Theorem \ref{thm5} implies that $n\leq q+t-1$.
\end{IEEEproof}

\begin{corollary}\label{shortOOA}
 For the $OOA(q^{t};t,m,t,q)$ given in Theorem~\ref{thm4} using $u=0$,
\[
m \leq \left\lfloor \dfrac{q}{\left\lfloor \frac{t}{2}\right\rfloor }\right\rfloor + 1.
\]
\end{corollary}

\begin{IEEEproof}
From Proposition~\ref{boundlinear}, we get $n\leq q+t-1$, which combined with
the relationship of $m$ and $n$ given in Theorem \ref{thm4} gives the desired bound  on $m$.
\end{IEEEproof}

In the next subsection, we look at Theorem \ref{thm4} again from a related but more abstract
point of view.

A well-known upper bound for the parameter $m$ from the theory of
digital nets over $\mathbb{F}_{q}$ is $m\leq \frac{q^{u+2}-1}{q-1}$.
Thus, for $u=0$, this upper bound is $m\leq q+1$.
Therefore, for $t\geq 4$ or for $t=3$ and $q$ odd, from the bound on $m$
given by Corollary~\ref{shortOOA} we conclude the construction
given in Theorem~\ref{thm4}
does not achieve the $q+1$ bound for digital nets.
In other words,  using $(n,t)$-sets all we can hope for are
linear OOAs bounded by
$ m \leq \left\lfloor \frac{q}{\left\lfloor \frac{t}{2}\right\rfloor }
     \right\rfloor + 1$, which is
generally much weaker than $q+1$.

On the other hand, our OOA construction in Theorem \ref{thm3} ({\it RUNS})
and the OOA construction in \cite{Skriganov,Rosenbloom} ({\it RTS})
build linear OOAs with $m=q+1$, which is the best possible.
In the rest of this section, we show that {\it RUNS}
is implicitly derived from a set of $n$ vectors in $\mathbb{F}_{q}^{t}$
such that not all choices of $t$ vectors are linearly independent
over $\mathbb{F}_{q}$. To the best of our knowledge this is a new
technique to construct OOAs, and we describe it as follows.

Let $f$ be a primitive polynomial of degree $t\geq 3$ over $\mathbb{F}_{q}$
and $\alpha \in \mathbb{F}_{q^{t}}$ be a root of $f$. Let
$S(f,T)=(a_{i})_{i\geq 0}$ be the LFSR sequence generated
by $f$ and $T=(b_{0},\ldots,b_{t-1})\in \mathbb{F}_{q}^{t}$,
$T\neq (0,\ldots, 0)$. By Proposition \ref{prop2}, there exists
a unique $\gamma \in \mathbb{F}_{q^{t}}$ such that
$T=(\mathrm{Tr}(\gamma\alpha^{0}), \ldots,
\mathrm{Tr}(\gamma\alpha^{t-1}))$. Since $\alpha$ is a
primitive element in $\mathbb{F}_{q^{t}}$, there exists
$v\in \{0,\ldots,q^{t}-2\}$ such that $\gamma=\alpha^{v}$, and so
$T=(\mathrm{Tr}(\alpha^{v}), \ldots, \mathrm{Tr}(\alpha^{v+t-1}))$.
 Let \linebreak $\Omega[q+1,t]$ be the set of labels of columns of $M$ given in Theorem \ref{thm3}.
For each $j\in \Omega[q+1,t]$ let $T_{j}=(b_{j,0},\ldots,b_{j,(t-1)})\in \mathbb{F}_{q}^{t}$
such that $b_{j,i}=\mathrm{Tr}(\alpha^{v+j}\alpha^{i})$ for $0\leq i \leq t-1$.
Consider the subset $\mathcal{X}$ of $\mathbb{F}_{q}^{t}$ given by
\[
\mathcal{X}=\{T_{j}\in \mathbb{F}_{q}^{t}: j\in \Omega[q+1,t]\}.
\]
The next result
gives a criterion to know which subsets of $\mathcal{X}$ of size $t$
are linearly independent over $\mathbb{F}_{q}$.

\begin{proposition}\label{prop6}
Under the conditions above, a subset $\{T_{j_{0}},\ldots,T_{j_{t-1}}\}$ of $\mathcal{X}$ is linearly
independent over $\mathbb{F}_{q}$ if and only if $\{\alpha^{j_{0}},\ldots,\alpha^{j_{t-1}}\}$
is linearly independent over $\mathbb{F}_{q}$.
\end{proposition}

\begin{IEEEproof}
By the conditions above,
$T_{j_{i}}=(\mathrm{Tr}(\alpha^{v+j_{i}}\alpha^{0}),\ldots,
\mathrm{Tr}(\alpha^{v+j_{j}}\alpha^{t-1}))$ for $0\leq i\leq t-1$.
Let $y_{0},\ldots,y_{t-1}\in \mathbb{F}_{q}$ not all of which are zero. The following relation holds
\setlength{\arraycolsep}{0.0em}
\begin{eqnarray}\label{eq5}
\sum_{i=0}^{t-1}y_{i}T_{j_{i}}&{}={}&\Bigg( \mathrm{Tr}\left(\alpha^{v}\sum_{i=0}^{t-1}y_{i} \alpha^{j_{i}}\right),\ldots, \nonumber\\
&&\mathrm{Tr} \left(\alpha^{v+t-1}\sum_{i=0}^{t-1}y_{i} \alpha^{j_{i}}\right)\Bigg).
\end{eqnarray}
\setlength{\arraycolsep}{5pt}
If $\{\alpha^{j_{0}},\ldots,\alpha^{j_{t-1}}\}$ is linearly dependent
over $\mathbb{F}_{q}$, then the set $\{T_{j_{0}},\ldots,T_{j_{t-1}}\}$
is linearly dependent over $\mathbb{F}_{q}$ by Eq.~(\ref{eq5}).
Conversely, if $\{T_{j_{0}},\ldots,T_{j_{t-1}}\}$ is linearly dependent
over $\mathbb{F}_{q}$ by Eq.~(\ref{eq5}), for $0\leq l \leq t-1$, we have
\[
\mathrm{Tr}\left(\alpha^{v+l}\left(y_{0} \alpha^{j_{0}}+\cdots
+y_{t-1} \alpha^{j_{t-1}}\right)\right)=0.\]
Since $\{\alpha^{v},\ldots,\alpha^{v+t-1}\}$ is linearly independent
over $\mathbb{F}_{q}$, we conclude that
\[
\mathrm{Tr}\left(\gamma\left(y_{0} \alpha^{j_{0}}+\cdots+
y_{t-1} \alpha^{j_{t-1}}\right)\right)=0,\]
for all $\gamma \in \mathbb{F}_{q^{t}}$. If
$b=y_{0} \alpha^{j_{0}}+\cdots+y_{t-1} \alpha^{j_{t-1}}$ is nonzero,
then $b$ has an inverse $b^{-1}\in \mathbb{F}_{q^{t}}$. For any
$\omega \in \mathbb{F}_{q^{t}}$, we can write $\omega=(\omega b^{-1})b$.
By choosing $\gamma=\omega b^{-1}$ we have
\[
0=\mathrm{Tr}\left(\gamma b \right)=\mathrm{Tr}\left(\left(\omega b^{-1}\right) b\right)=\mathrm{Tr}\left(\omega\right).
\]
This implies that $\mathrm{Tr}\left(\omega \right)=0$ for all $\omega \in \mathbb{F}_{q^{t}}$,
which is not possible. Therefore $b=0$, implying that
$\{\alpha^{j_{0}},\ldots,\alpha^{j_{t-1}}\}$ is linearly dependent over $\mathbb{F}_{q}$.
\end{IEEEproof}

Now we see how the previous result relates to our OOA
construction. For each $T_{j} \in \mathcal{X}$, let
$S(f,T_{j})$ be the LFSR sequence generated by $f$ and
$T_{j}$. Consider the array whose columns are
$C^{q^t-1}_0(S(f,T_{j}))$ for $j \in \Omega[q+1,t]$.
Add to this array the all-zero row. This array is exactly
the OOA constructed in Theorem \ref{thm3}, thus these
OOAs are constructed  from the set $\mathcal{X}$ of
vectors in $\mathbb{F}_{q}^{t}$. The set  $\mathcal{X}$
is not an $(|\mathcal{X}|,t)$-set in $\mathbb{F}_{q}^{t}$,
since $|\mathcal{X}|\geq q+t$. Indeed, for every
$K\subseteq\mathcal{X}$ with $|K|\geq q+t$, we conclude
$K$ is not a $(|K|,t)$-set in $\mathbb{F}_{q}^{t}$,
otherwise by Proposition~\ref{boundlinear} we could
construct an $OA(q^{t};t,|K|,q)$.

In the next subsection, we generalize the notion of linear independence using
hypergraphs and see that Theorem \ref{thm4} and our Theorem \ref{thm3} belong to a
family of constructions which are instances of a general principle given
in Theorem \ref{thmraph}.

\subsection{Ordered orthogonal arrays, hypergraphs and homomorphisms}\label{sec62}

The construction in this paper can be viewed through  the lens of hypergraph homomorphisms, which has the potential for further fruitful use in the construction of ordered orthogonal arrays and of other generalizations of orthogonal arrays.
In some sense, we are abstracting the notion of linear independence with hypergraphs and
abstracting the notion of the choices of columns with homomorphisms.
Informally to start, suppose we have an $N \times k$ $v$-ary array that has a collection $H$ of $t$-subsets of the columns that is $\lambda$-covered.  In the previous subsection,
the array was either an OA and $H$ the set of all $t$-sets of columns or the
array was a subinterval array from an LFSR sequence and $H$ the set of all linearly
independent $t$-sets of columns.
Considering $H$ as a hypergraph, suppose we additionally have a homomorphism from another hypergraph $G \rightarrow H$.  Then we can construct an $N \times |V(G)|$ $v$-ary  that is  $\lambda$-covered on the hyperedges of $G$.
This map simply records the choices of columns from one array we are using to
build a new array as described by Theorems \ref{thm4} and \ref{thm3}, respectively.
We now make this construction formal and review  existing results on $(t,m,s)$-net and ordered orthogonal array literature, including our construction,  via this framework.

A $t$-uniform {\em hypergraph}, $G$, consists of a finite vertex set $V(G)$ and a collection, $E(G)$, of $t$-subsets of $V(G)$ called {\em hyperedges}.  Let $H_{t,m,s}$ be the hypergraph whose hyperedges are precisely the left-justified subsets of size $t$ of $[m]\times [s]$.  A {\em complete $t$-uniform hypergraph of order $n$}, $K_{n}^{t}$, is the hypergraph $(X,E)$ where $|X|=n$ and $E = {X \choose t}$, where ${X \choose t}$ is
the set of all subsets of $X$ of cardinality $t$.   Given a finite vector space, $\mathbb{F}_{q}^d$, the {\em linear independence hypergraph}, $LI_{d,q}$ is the hypergraph with $V(LI_{d,q}) = \mathbb{F}_{q}^{d}$ and $e=\{v_0,v_1, \ldots, v_{d-1}\} \in E(LI_{d,q})$ if and only if $e$ is a linearly independent set over $\mathbb{F}_{q}$ (equivalently $e$ is not contained in any dimension $d-1$ subspace).  Similarly given a projective space $PG(d,q)$, the {\em projective independence hypergraph}, $PI_{d,q}$ is the hypergraph with the $(q^{d+1}-1)/(q-1)$ points of the space as vertices and $e=\{v_0,v_1, \ldots, v_{d}\} \in E(PI_{d,q})$ if and only if $e$ is not contained in any dimension $d-1$ subspace.  A reference
for the geometry concepts used here is \cite{cameron}.

\begin{definition}
For two $t$-uniform hypergraphs $G,H$, a {\em hypergraph homomorphism} $f\colon G \rightarrow H$ is a map \linebreak $f\colon V(G) \rightarrow V(H)$ such that if $e = \{v_0, \ldots, v_{m-1}\} \in E(G)$, then $f(e) = \{f(v_0), \ldots, f(v_{m-1})\} \in E(H)$, and $|e| = |f(e)|$.
\end{definition}

The fact that there exists an injective homomorphism \cite{cameron}
\[
PI_{d,q} \rightarrow LI_{d+1,q}
\]
gives a geometrical motivation for the truncation of subintervals of the LFSR at length $(q^t-1)/(q-1)$ used in Section \ref{sec4}.

\begin{definition}
Let $\mathcal{G}$ be a $t$-uniform hypergraph on $k$ vertices. A {\em variable strength orthogonal array}, $VOA(N; \mathcal{G}, v)$, is an $N \times k$ array over $\{0, \ldots, v-1\}$ with columns labeled by $V(\mathcal{G})$ such that if $B = \{b_0, \ldots, b_{t-1}\} \in E(\mathcal{G})$, then the $N \times t$ subarray labeled by $B$  is $\lambda$-covered, where $\lambda=N/v^t$.
\end{definition}

In this language, an $OOA_{\lambda}(N;t,m,s,v)$ is equivalent to a $VOA(N;H_{t,m,s},v)$, and an $OA_{\lambda}(N;t,n,v)$
is a $VOA(N; K_{n}^{t},v)$.

The following theorem is a general purpose construction linking homomorphisms and variable strength orthogonal arrays.

\begin{theorem}\cite{RaaphorstT}\label{thmraph}
Let $G$ and $H$ be $t$-uniform hypergraphs.  Suppose that there exists a $VOA(N;H,v)$ and a hypergraph homomorphism $f\colon G \rightarrow H$.  Then there exists a $VOA(N;G,v)$.
\end{theorem}

For this construction we need two things:  an already existing array where the edges of the hypergraph $H$ describes the sets of columns that are
$\lambda$-covered, and a homomorphism from hypergraph $G$ to $H$.  In most of the uses of this construction that we are aware of, the hypergraph $H$ is well known and the work is establishing the homomorphism.   This framework also applies to more general objects called
covering arrays \cite{RaaphorstT}.

We now demonstrate how previous results on $(t,m,s)$-nets and OOAs can be viewed with the hypergraph homomorphism language.
 Theorem~2 of \cite{mr1169232}, Theorem~5.4 of \cite{N},  and Theorem~1 of \cite{mr1181933} all use the existence of an $OA_{\lambda}(N;2,m,v)$, equivalently a $VOA(N;K_{m}^2,v)$, to construct an $OOA_{\lambda}(N;2,m,2,v)$ by establishing a homomorphism from $H_{2,m,2} \rightarrow H_{2,m,1} = K_{m}^{2}$.  In homomorphism terms, they also use the fact that there is a more obvious homomorphism $K_{m}^{2} \rightarrow H_{2,m,2}$ to show that the existence of $OA_{\lambda}(N;2,m,v) = OOA_{\lambda}(N;2,m,1,v)$ is equivalent to the existence of an  $OOA_{\lambda}(N;2,m,2,v)$. Theorem~3 in \cite{Mullen} extends this to higher strength showing that the existence of an $OOA_{\lambda}(N;t,m,t-1,v)$ is equivalent to an $OOA_{\lambda}(N;t,m,t,v)$ by establishing the homomorphisms
\[
H_{t,m,t-1} \rightarrow H_{t,m,t} \rightarrow H_{t,m,t-1}.
\]
They point out that this yields a construction when the $OOA_{\lambda}(N;t,m,t-1,v)$ is known to exist which is powerful in the $t=2$ case, since there are many known $OA_{\lambda}(N;2,m,v) = OOA_{\lambda}(N;2,m,1,v) = VOA(N;K_{m}^{2},v)$. In some of these cases the OOAs are linear and the $VOA(N;K_{m}^{2},q)$ is derived from $(n,t)$-sets and thus linear independence plays a role in the constructions.

The Hammersley net, an $OOA(t,2,t,v)$ given in Example~14.8.5 in \cite{HFF}, is constructed by giving a homomorphism from $H_{t,2,t} \rightarrow K_{t}^{t} = H_{t,1,t}=H_{t,t,1}$ and using the trivial $OOA(t,t,1,v) = OA(t,t,v) = VOA(v^{t};K_{t}^{t},v)$.  Example~14.8.6 in \cite{HFF} constructs an $OOA_{\lambda}(N;1,m,1,v)$ by essentially using a homomorphism from $K_{m}^{1} \rightarrow K_{1}^{1}$.

Theorem~5.1 in \cite{Lawrence} and Theorem~1 in \cite{Maha} use the existence of an
$OA_{\lambda}(N;t,n,v)$ to construct an $OOA_{\lambda}(N;t,m,t,v)$ where
if $t=2h+\delta$, $\delta \in \{0,1\}$, we have $n=mh+\delta$.
From the hypergraph homomorphism point of view, both articles can be seen as establishing a homomorphism from \linebreak $H_{t,m,t} \rightarrow K_{n}^{t}$.
This is the equivalent of Theorem \ref{thm4} from the previous subsection.

Fuji-Hara and Miao \cite{fuji-hara} determine the geometric structures in $PG(d,q)$ that are equivalent to linear $OOA_{q^{d-2}}(q^{d+1};3,m,3,q)$ and  $OOA_{q^{d-3}}(q^{d+1};4,m,4,q)$.  This can be viewed as determining the existence of homomorphisms
\[
H_{3,m,3} \rightarrow PI_{d,q}
\]
for prime powers $q$, when $q=2$ and $m \leq 2^{d}-1$ or when $q > 2$ and there exists a set of $m$ points in $PG(d,q)$, no three of which are co-linear; and
\[
H_{4,m,4} \rightarrow PI_{d,q}
\]
for prime powers $q$, when a particular point configuration exists in $PG(d,q)$.  They then show that this configuration always exists when $d=3$ and $m=q+1$.  In some of their methods, there are non-constructive proofs for the point configurations.

In the hypergraph homomorphism language, our Theorem~\ref{thm3} is proved by using the existence of the subinterval array from the LFSR which is a $VOA(q^t;PI_{t-1,q},q)$ and then building a homomorphism
\[
H_{t,q+1,t} \rightarrow PI_{t-1,q}.
\]

All these examples show the power of the hypergraph homomorphism
technique to construct OOAs.
As pointed out by Martin \cite{bill} in discussions with the fourth author,
the existing OOA constructions prior to 2002 (all the work surveyed above
prior to our own and that of Fuji-Hara and Miao)  essentially repeated columns
of existing orthogonal arrays in clever ways so that the
resulting arrays satisfied the required column coverage for
the OOA definition.  Martin \cite{bill} conjectures that there are similar
homomorphic techniques that construct OOAs, but which do not
simply repeat columns.  The work of Fuji-Hara and
Miao for $t=3,4$ and our Theorem~\ref{thm3} for arbitrary $t$
are the first such homomorphic constructions of OOAs which meet
this goal. Our work is fully constructive.

Another use of hypergraph homomorphisms in the OOA and
$(t,m,s)$-net literature is to prove non-existence results.
Because there is a simple homomorphism
\[
K_{m}^{t} \rightarrow H_{t,m,t},
\]
Theorem~\ref{thmraph} shows that if an $OA_{\lambda}(N;t,m,v)$ does not
exist then an $OOA_{\lambda}(N;t,m,t,v)$ cannot exist. This homomorphism
technique is the essence of several non-existence results for
OOAs in the literature \cite{N}, \cite{Maha}.  Fuji-Hara and Miao
\cite{fuji-hara}, in homomorphism terms, use this to prove that an \linebreak
$OOA_{q^{d-4}}(q^d;4,m,4,q)$ cannot exist unless
\[
(q+1)m + (q-1){m \choose 2} \leq \frac{q^{d}-1}{q-1}
\]
and
\[
m \leq \frac{q^d-2}{q-1}.
\]
Since this method is so general, we believe that it has the potential to yield new non-existence results.
The goal should be to find a hypergraph homomorphism $G \rightarrow H_{t,m,s}$ and also show that a $VOA(N;G,v)$ does not exist.
This would establish a non-existence result for $OOA_{\lambda}(N;t,m,s,v)$.

We also believe that the homomorphism construction has great
potential to yield new constructions of ordered orthogonal
arrays, variable strength orthogonal arrays and variable
strength covering arrays.  In particular, in Section~\ref{sec5},
for $3\leq t \leq 5$, we experimentally counted the number of
$t$-sets of columns that were covered in our construction of
Theorem~\ref{thm3} and also in the Rosenbloom-Tsfasman-Skriganov
construction \cite{Rosenbloom,Skriganov}.  It would be
interesting to characterize the full set of hyperedges that
are covered in the arrays from these constructions.

\section*{Acknowledgment}
Brett Stevens would like to thank William J. Martin for his
many discussions over the years about the construction of
orthogonal array-like objects using hypergraph homomorphisms.

The authors would like to thank the anonymous referees for their suggestions that
greatly improved this paper.

\ifCLASSOPTIONcaptionsoff
  \newpage
\fi




\begin{thebibliography}{1}



\bibitem{cameron} P.~J.~Cameron. (2000) Projective and Polar Spaces. [Online]. Available:
\url{http://www.maths.qmul.ac.uk/~pjc/pps/}.

\bibitem{Colburn} C.~J.~Colbourn and J.~H.~Dinitz, \emph{Handbook of
Combinatorial Designs}. Second Edition, Chapman $\&$ Hall/CRC, Boca Raton, 2006.


\bibitem{Dewar} M.~Dewar, L.~Moura, D.~Panario, B.~Stevens, and Q.~Wang,
``Division of trinomials by pentanomials and orthogonal arrays,''
\emph{Des. Codes Cryptogr.}, vol.~45, pp.~1--17, 2007.

\bibitem{fuji-hara} R.~Fuji-Hara and Y.~Miao, ``A note on geometric
structures of linear ordered orthogonal arrays and $(t,m,s)$-nets
of low strength,'' \emph{Des. Codes Cryptogr.}, vol.~26, pp.~257--263, 2002.

\bibitem{Golomb} S.~W.~Golomb and G.~Gong, \emph{Signal Design for Good
Correlation for Wireless Communication, Cryptography, and Radar}.
Cambridge University Press, Cambridge, 2005.

\bibitem{Hedayat} A.~S.~Hedayat, N.~J.~A.~Sloane, and J.~Stufken,
\emph{Orthogonal Arrays: Theory and Applications}. Springer, 1999.


\bibitem{Tamar} T.~Krikorian, \emph{Combinatorial Constructions of
Ordered Orthogonal Arrays and Ordered Covering Arrays}. MSc.~Thesis,
Ryerson University, 2011.

\bibitem{Lawrence} K.~M.~Lawrence, ``A combinatorial characterization
of $(t,m,s)$-nets in base $b$,'' \emph{J. Combin. Des.}, vol.~4, pp.~275--293, 1996.

\bibitem{Maha} M.~Lawrence, A.~Mahalanabis, G.~L.~Mullen, and W.~Ch.~Schmid, ``Construction of digital $(t,m,s)$-nets from
linear codes'' in: S.~D.~Cohen and H.~Niederreiter (eds.), \emph{Finite
Fields and Applications}, London Math.~Soc.~Lecture
Note Ser.~233, pp.~189--208, 1996.

\bibitem{Lidl} R.~Lidl and H.~Niederreiter,  \emph{Finite Fields}.
Cambridge University Press, Cambridge, 1997.

\bibitem{bill} W.~J.~Martin, personal communication.

\bibitem{DCCsurvey} L.~Moura, G.~L.~Mullen, and D.~Panario, ``Finite field constructions of
combinatorial arrays,'' \emph{Des. Codes Cryptogr}. vol.~78, pp.~197--219, 2016.

\bibitem{HFF} G.~L.~Mullen and D.~Panario, \emph{Handbook of Finite Fields}.
Chapman $\&$ Hall/CRC, Boca Raton, 2013.

\bibitem{Mullen}  G.~L.~Mullen and W.~Ch.~Schmid, ``An equivalence
between $(t,m,s)$-nets and strongly orthogonal hypercubes,''
\emph{J. Combin. Theory A}, vol.~76, pp.~164--174, 1996.

\bibitem{mr1169232} G.~L.~Mullen and G.~Whittle, ``Point sets with
uniformity properties and orthogonal hypercubes,''
\emph{Monatsh. Math.}, vol.~113, pp.~265--273, 1992.

\bibitem{Munemasa} A.~Munemasa, ``Orthogonal arrays, primitive trinomials,
and shift-register sequences,'' \emph{Finite Fields Appl.}, vol.~4, pp.~252--260, 1998.

\bibitem{N} H.~Niederreiter, ``Point sets and sequences with small
discrepancy,'' \emph{Monatsh. Math.}, vol.~104, pp.~273--377, 1987.

\bibitem{NiederMC} H.~Niederreiter,  \emph{Random Number Generation and
Quasi-Monte Carlo Methods}. CBMS-NSF Series in Applied Math.,
SIAM, Philadelphia, 1992.

\bibitem{mr1181933} H.~Niederreiter, ``Orthogonal arrays and other
combinatorial aspects in the theory of uniform point distributions
in unit cubes,'' \emph{Discrete Math.}, vol.~106--107, pp.~361--367, 1992.

\bibitem{Panario} D.~Panario, O.~Sosnovski, B.~Stevens, and Q.~Wang,
``Divisibility of polynomials over finite fields and combinatorial
applications,'' \emph{Des. Codes Cryptogr.}, vol.~63, pp.~425--445, 2012.

\bibitem{RaaphorstT} S.~Raaphorst, \emph{Variable Strength Covering Arrays}.
Ph.~D.~Thesis, University of Ottawa, 2013.

\bibitem{Raaphorst} S.~Raaphorst, L.~Moura, and B.~Stevens, ``A
construction for strength-3 covering arrays from linear feedback
shift register sequences,'' \emph{Des. Codes Cryptogr.},  vol.~73, pp.~949--968, 2014.

\bibitem{Roman} S.~Roman, \emph{Coding and Information Theory}. Springer, 1992.

\bibitem{Rosenbloom}  M.~Y.~Rosenbloom and M.~A.~Tsfasman,  ``Codes for
$m$-metric,'' \emph{Probl. Inf. Transm.}, vol.~33, pp.~45--52, 1997.

\bibitem{Skriganov} M.~M.~Skriganov, ``Coding theory and uniform
distributions,`` \emph{St. Petersburg Math. J.}, vol.~33, pp.~301--337, 2002.

\bibitem{Tzanakis} G.~Tzanakis, L.~Moura, D.~Panario, and B.~Stevens,
``Constructing new covering arrays from LFSR sequences over finite
fields,`` \emph{Discrete Math.}, vol.~339, pp.~1158--1171, 2016.

\end{thebibliography}
%

\newpage

\begin{IEEEbiographynophoto}{Andr\'{e} Guerino Castoldi}
was born in Francisco Beltr\~{a}o, Brazil, in 1988. He graduated in Mathematics (2010)
from the Federal Technological University of Paran\'{a} (UTFPR) in Pato Branco, Brazil. He received
the M.Sc. and Ph.D. degrees in Mathematics from the State University of Maring\'{a} (UEM), Brazil,
in 2012 and 2016, respectively. From September 2014 to August 2015, he was a visiting researcher
at the University of Ottawa, Canada, supported by the Science without Borders Program, Brazil.
Currently, he is an Assistant Professor at the Federal Technological University of Paran\'{a} (UTFPR) in Pato Branco, Brazil.
His research interests are combinatorial designs, covering codes, poset metrics, finite fields and their relationships.
\end{IEEEbiographynophoto}

\begin{IEEEbiographynophoto}{Lucia Moura}
 received her B.Sc. in Computer Science (1987) and M.Sc. in Applied
Mathematics (1992) from the University of S\~ao Paulo, Brazil. In 1999,
she received her Ph.D. in Computer Science from the University of Toronto,
Canada. She spent one year as a postdoctoral fellow at the Fields
Institute for Research in Mathematical Sciences in Toronto and then joined
the School of Electrical Engineering and Computer Science, University of
Ottawa, where she is currently an Associate Professor. Her research
interests include combinatorial algorithms, combinatorial designs and
their applications.
\end{IEEEbiographynophoto}

\begin{IEEEbiographynophoto}{Daniel Panario}
was born in Uruguay where he studied Mathematics and
Computer Science. He received a M.Sc. degree from the University of
S\~ao Paulo (Brazil) and a Ph.D. from the University of Toronto
(Canada). He is a Professor at Carleton University in Ottawa (Canada),
and a Senior Member of IEEE. His main research interests are in
finite fields and their applications in information theory and
communications, as well as in the analysis of algorithms.
\end{IEEEbiographynophoto}

\begin{IEEEbiographynophoto}{Brett Stevens}
was educated at the University of Chicago, University College
London, and the University of Toronto. His M.Sc. was in mathematical biology
and his Ph.D. in mathematics, specifically combinatorics. He did post-doctoral
work at Simon Fraser University and IBM T.~J.~Watson Laboratories. He is
interested in combinatorics, applications of mathematics and the interaction of
mathematics with other disciplines and culture. He is a Professor in Mathematics
at Carleton University.
\end{IEEEbiographynophoto}

\end{document}